\documentclass[12pt,authoryear]{article}
\usepackage[english]{babel}
\usepackage{comment}
\usepackage[letterpaper,top=2cm,bottom=2cm,left=3cm,right=3cm,marginparwidth=1.75cm]{geometry}

\usepackage{graphicx}
\usepackage[colorlinks=true, allcolors=blue]{hyperref}
\usepackage[font=footnotesize]{caption}
\usepackage[textwidth=8em,textsize=small]{todonotes}
\usepackage{color}
\usepackage{amsfonts}
\usepackage{amssymb,bm}
\usepackage{enumerate,rotate}
\usepackage{amssymb,latexsym,mathrsfs}
\usepackage[noend]{algpseudocode}
\usepackage{algorithm}
\algnewcommand{\IfThenElse}[3]{
  \State \algorithmicif\ #1\ \algorithmicthen\ #2\ \algorithmicelse\ #3}
\usepackage{graphicx}
\usepackage[toc,page]{appendix} 
\usepackage{natbib}
\RequirePackage[OT1]{fontenc}
\RequirePackage{amsthm,amsmath}
\usepackage{authblk}

\newcommand{\E}{\mathbb{E}}

\def\I{\mbox{I}}

\newcommand{\N}{\mathds{N}}

\def\E{\mbox{E}}

\def\N{\mathcal{N}}

\numberwithin{equation}{section}
\theoremstyle{plain}

\theoremstyle{definition}

\theoremstyle{remark}

\title{Fast Bayesian inference in a class of sparse linear mixed effects models}
\author[1]{M-Z. Spyropoulou}
\author[1]{J. Hopker}
\author[2]{J. E. Griffin}
\affil[1]{School of Sport and Exercise Sciences, University of Kent, U. K.}
\affil[2]{Department of Statistical Science, 
University College London, Gower Street, 
London WC1E 6BT,
U. K., email: j.griffin@ucl.ac.uk}
\date{}

\begin{document}
\maketitle
\begin{abstract}
Linear mixed effects models are widely used in statistical modelling. 
We consider a mixed effects model with Bayesian variable selection in the random effects using spike-and-slab priors and developed a variational Bayes inference scheme that can be applied to large data sets. An EM algorithm is proposed for the model with normal errors where  
the posterior distribution of the variable inclusion parameters is approximated using an Occam's window approach. Placing this approach within a variational Bayes scheme also the algorithm to be 
 extended to the model with skew-$t$ errors. The performance of the algorithm is evaluated in a simulation study and applied to a 
 longitudinal model for elite athlete performance in the 100 metre sprint and weightlifting. 
\end{abstract}

\section{Introduction}

Linear Mixed Effects (LME) models are widely used when we have multiple observations on a sample of individuals, such as 
repeated measures \citep[e.g.][]{lindsey99}, longitudinal measurements \citep[e.g.][]{fitz08} or semiparametric regression models \citep[e.g.][]{rupp03}. Suppose that 
 there are  $M$ individuals with the $i$-th individual having observations
 $\boldsymbol{y}_i = (y_{i, 1}, \dots, y_{i, n_i})$, and design matrices 
$\boldsymbol{X}_{i}$ ($n_i\times q$) and $\boldsymbol{S}_{i}$ ($n_i\times p$) 
  modelled by
\begin{equation}
\boldsymbol{y}_{i} = \zeta_0 + \boldsymbol{X}_{i}\,\boldsymbol{\zeta}  + \boldsymbol{S}_{i}\,\boldsymbol{\beta}_i 
+ \boldsymbol{\epsilon}_i 
\label{lme}
\end{equation}
where $\zeta_0$ and $\boldsymbol\zeta$ are fixed effects whose values are shared by all individuals, $\boldsymbol{\beta}_i$ are individual-specific zero-mean random effects with covariance matrix $\boldsymbol\Omega$, and $\boldsymbol{\epsilon}_i = (\epsilon_{i, 1}, \dots, \epsilon_{i, n_i})$ are i.i.d. zero-mean errors. In a Bayesian analysis of this model,
the random effects and the errors are usually assumed to be normally distributed.

In modern applications, either $p$ or $q$ (or both) may be high-dimensional which has led to the use of variable selection in LME models. A prior distribution is given to $\boldsymbol\zeta$ and/or $\boldsymbol\Omega$ which encourages elements to be shrunk towards zero. 
Initial methodological and computational developments concentrated  on the application of  variable selection to both the fixed and random effects by extending methods developed for linear models. 
\cite{ChenDunson} suggested using 
the form $\boldsymbol\Omega = \mbox{diag}(\boldsymbol\lambda) \,\boldsymbol{B} \,\mbox{diag}(\boldsymbol\lambda)$ where $\boldsymbol{B}$ is a $(p\times p)$-dimensional matrix and $\boldsymbol\lambda$ is $p$-dimensional vector.  Spike-and-slab  variable selection priors are placed on $\boldsymbol\lambda$ and $\boldsymbol\zeta$ and inference uses  Markov chain Monte Carlo (MCMC) methods. This approach was subsequently developed to use variational Bayes inference with shrinkage priors by \cite{ArmaganDunson}.
More recently, work has focused on variable selection in the fixed effects only (and so the random effects are effectively treated as nuisance parameters) using variational Bayes methods.
\cite{Tung19} consider using a Bayesian lasso prior
\citep{ParCas08}
for the fixed effects. This approach is extended by \cite{Degani22} to allow for general global-local priors \citep{BhaDat19},
 more sophisticated random effect structures and to take advantages of fast matrix methods.

In contrast to previous work, we focus on Bayesian variable selection in the setting where $p$ is high-dimensional, $n_i$ can be large and  the covariance of the random effects $\boldsymbol\Omega$ is assumed to be diagonal. This allows us to consider variable selection at the individual level with potentially different variables used as random effects for each individual. This set-up is motivated by recent work in modelling elite sporting performance in events such as 100 metres sprints or weightlifting over an athlete's career. Interest focuses on the trajectory of an individual's sporting performance as a function of age \cite[see {\it e.g.}][]{berry1999bridging} and these differ between individuals due to individual physiology, injuries, training, etc. 
We consider the approach of 
\cite{griffin2022bayesian} who use a linear mixed effects model where the fixed effects include polynomial terms for age (providing a population effect of age), as well as  environmental conditions (such as wind speed), the month of the event, or the prestige of an event. The difference between an individual's trajectory and the population effect of age is modelled using linear splines as random effects.
Their model is applied to large databases containing thousands of athletes with potentially hundreds of performances. 

The large  number of variables, {\it i.e.} splines, in the individual's age trajectories motivates the use of variable selection and the large number of observations per individual allow us to have individual variable selection. \cite{griffin2022bayesian} use MCMC to fit the model but this can be slow and involve substantial amounts of memory. The novelty and contribution of the paper is to 
develop an EM-based \citep{DemLaiRub77, MengDyk97} and variational Bayes approach \citep{BleiKucMcA17} for  Bayesian inference in this model with both normal and skew $t$ errors. 
We approximate the posterior distribution on the individual variable inclusion parameters using an Occam's window approach  \citep{madigan1994model} and show how this can be included in an EM-type and variational Bayes algorithms.

The paper is organized as follows: Section \ref{model} explains how the model is formed. In Section \ref{infer}, an EM algorithm for inference in the LME model with normal errors is developed including the approximation of the posterior distribution of the variable inclusion indicators using the 
Occam's window approach.
  In section \ref{t-skew} an extension to non-normal error is presented which uses the Occam's window approximation in a variational Bayes algorithm and is developed for the specific case of  skew $t$ errors. Section \ref{results} includes a simulation study and 
applications to 100 metre sprinting and weightlifting data with  
a comparison of the algorithm to the MCMC algorithm in \cite{griffin2022bayesian}. Lastly, Section \ref{conclusions} concludes. Appendices gives further details of the algorithms and further results from the simulation study.

\section{Sparse Linear Mixed Effects Models}\label{model}

 We consider the LME model in \eqref{lme} and  initially assume that $\epsilon_{i,j}\stackrel{ind.}{\sim}\N\left(0, \sigma^2_i\right)$ (we will consider relaxing this assumption to a skew $t$ error distribution in Section \ref{t-skew}).
  The Bayesian variable selection approach introduces indicator variables 
   $\boldsymbol\gamma_i = (\gamma_{i,1},\dots, \gamma_{i, p})$
  for the $i$-th individual 
  where $\gamma_{i, j}$ is 1 if the $j$-th random effect is included in the model and is 0 otherwise.
  We write $\boldsymbol\zeta^{\star}$ = $(\zeta_0, \boldsymbol\zeta)$,
  $\boldsymbol{S}_i^\gamma$ as the design matrix including only random effects with $\gamma_{i, j} = 1$,
$p_i^{\gamma} = \sum_{j=1}^p \gamma_{i, j}$ for the number of selected random effects
and $\boldsymbol{\beta}^{\gamma}_i$ for the corresponding coefficients. The LME model becomes
\begin{equation*}
\boldsymbol{y}_{i} = \boldsymbol{X}_{i}\,\boldsymbol{\zeta}^*  + \boldsymbol{S}^{\gamma}_{i}\,\boldsymbol{\beta}^{\gamma}_i+ \boldsymbol{\epsilon}_i.
\end{equation*}
The Bayesian model is completed by assigning priors to the parameters. The commonly-used beta-binomial prior is used for the inclusion variables $\boldsymbol\gamma_i$ so that 
$\gamma_{i,k} \stackrel{i.i.d.}{\sim} \mathcal{B}ernoulli (h_i)$
and $ h_i \sim \mathcal{B}e \left(a_1,b_1\right)$. This implies that
\[
p(\boldsymbol\gamma_{i}) = \frac{\Gamma(a_1 + b_1)}{\Gamma(a_1)\Gamma(b_1)}
\frac{\Gamma( p_i^{\gamma} + a_1 )\Gamma( p - p_i^{\gamma}+ b_1)}{\Gamma(p+ a_1 + b_1 )}.
\]
The regression coefficients for the included variables are
$\beta^{\gamma}_{i,1}\sim\N(0, \psi\sigma^2_i)$ and $\beta^{\gamma}_{i, j}\mid \gamma_i \sim \N(0, g^2\sigma^2_i)$ for $j=2,\dots, p^{\gamma}_i+1$. We assume a vague prior for the fixed effects, $p(\boldsymbol\zeta^*)\propto 1$ but other choices such as a normal prior, $g$-prior, or global-local shrinkage prior could be used. 
The error variance is assumed to be $\sigma^{2}_i \stackrel{i.i.d.}{\sim} \mathcal{I}G(a, b)$. The prior is completed by specifying the hyperpriors:
$\psi \sim \mathcal{I}G(1,1)$, where $\mathcal{I}G(a,b)$ represents an inverse-gamma distribution with shape $a$ and mean $b/(a-1)$ if $a>1$, $\sqrt{g} \sim \mathcal{HC}(1)$, where $\mathcal{HC}(\gamma)$ represents a half-Cauchy distribution with scale $\gamma$, $p(a_1, b_1) \propto 1$ and $p(a, b) \propto 1$. 

\subsubsection*{Example: Athletic Performance}

The LME model for elite sporting performance in \cite{griffin2022bayesian} assumes that
  $y_{i, j}$ is the $j$-th performance of the $i$-th athlete. The fixed effects $\boldsymbol{X}_{i}$ are a column of ones for the intercept $\zeta_0$, powers of age (\cite{griffin2022bayesian} use a fourth-order polynomial) which leads to a flexible \textit{population-level} age effect and confounders (for example, wind speed in 100 metres sprints or pool size in swimmming events).
    A polynomial is sufficient to model the population effect of age but not the \textit{individual-level age}
 effect (which can include changes in performance due to injury, loss of form or changes in training). A piecewise linear spline is a more flexible form and leads to
 an $(n_{i} \times p)$-dimensional design matrix
 $\boldsymbol{S}_{i}$  of spline basis functions evaluated at the observed ages. A large value of $p$ is chosen to allow flexibility with the Bayesian variable selection prior used to avoid overfitting.

\section{EM algorithm with normal errors}\label{infer}

We develop an approach to inference in the model in \eqref{lme} when the sample size $M$ is large. The parameters are divided into a set of population-level parameters $\boldsymbol{\chi} = \{\zeta_0, \boldsymbol\zeta, \psi, g^2, a, b, a_1, b_1   \}$ and a set of 
    individual-level parameters $\boldsymbol{\nu}_i = \{\boldsymbol\gamma_i, \,
    \boldsymbol\beta_i, \,\sigma_i^2\}$. The maximum a posterior (MAP) estimate  of 
 $\boldsymbol{\chi}$ integrating over the individual-level parameters 
  $\boldsymbol{\nu}_1, \dots,  \boldsymbol{\nu}_M$ is found using the EM algorithm \citep{DemLaiRub77, MengDyk97}.
 To simplify the notation, we drop the $\gamma$ superscript.
 The EM algorithm iterates the following two steps:
\begin{enumerate}
\item {\bf Expectation Step}: Calculate  
\begin{align}
Q(\boldsymbol{\chi}) =& 
\sum_{i=1}^M \E_{\boldsymbol{\nu}_i\mid \boldsymbol{\chi},\boldsymbol{y}_i} \left[ \log P\left(\boldsymbol{y}_i\mid \boldsymbol{\nu}_i, \boldsymbol{\chi} \right) + \log P\left(\boldsymbol{\nu}_i|\boldsymbol{\chi} \right)\right]  + \log P\left(\boldsymbol{\chi}\right) 
\notag
\\
=& \sum_{i=1}^M 
\E_{\boldsymbol\gamma_i\mid \boldsymbol{\chi}, \boldsymbol{y}_i}\left[
\E_{\boldsymbol\beta_i, \sigma_i^2 \mid \boldsymbol\gamma_{i}, \boldsymbol{\chi}, \boldsymbol{y}_i}
\left[ \log P\left(\boldsymbol{y}_i\mid \boldsymbol\gamma_i, \,
    \boldsymbol\beta_i, \, \sigma_i^2,\boldsymbol{\chi} \right)+ \log P(\boldsymbol\gamma_i, \,
    \boldsymbol\beta_i, \, \sigma_i^2\mid\boldsymbol\chi) 
\right]
\right]\notag\\
&+\log P\left(\boldsymbol{\chi}\right) 
\label{simple_EM}
\end{align}
\item {\bf Maximization Step}: Find $\text{arg max}_{\boldsymbol{\chi}} Q(\boldsymbol{\chi})$
\end{enumerate}

To use the EM algorithm we need  to calculate the posterior distributions of the individual-specific parameters $\boldsymbol{\nu}_1, \dots, \boldsymbol{\nu}_M$. Importantly, 
$\boldsymbol{\nu}_1,\dots,\boldsymbol{\nu}_M$  are conditionally independent given $\boldsymbol{\chi}$ under the posterior distribution. A challenging aspect of using the algorithm is that the expectation with respect to $\boldsymbol\gamma_i$  involves a sum
over the model space parameter $\boldsymbol\gamma_i$ which is discrete with $2^p$ possible values. The computational time needed to evaluate this sum increasingly exponentially with $p$ and the sum cannot be fully enumerated for $p$ greater than 30. We  use
 the  Occam's window approach \citep{madigan1994model} to
approximate this sum by a sum of $K\ll 2^p$ models. These models will be denoted $\boldsymbol\gamma_{i, 1}, \dots, \boldsymbol\gamma_{i, K}$ with associated parameters $(\boldsymbol{\beta}_{i, 1}, \sigma_{i, 1}^2), \dots, (\boldsymbol{\beta}_{i, K}, \sigma_{i, K}^2)$. The expectation of a function $f(\boldsymbol\gamma_i)$ is approximated by the sum 
\begin{align}
\sum_{k=1}^K w_{i, k} \, f\left(\boldsymbol{\gamma}_{i, k}\right)
\label{approx1}
\end{align}
where
\begin{align}
 w_{i,k}= \frac{p\left(\boldsymbol\gamma_{i, k}\mid \boldsymbol{\chi} \right) \,m_i\left(\boldsymbol\gamma_{i, k}\right)}{\sum_{t=1}^K p\left(\boldsymbol\gamma_{i, t}\mid  \boldsymbol{\chi} \right) \, m_i\left(\boldsymbol\gamma_{i, t}\right)}
\label{weight}
\end{align} 
and
\begin{align*}
m_i\left(\boldsymbol\gamma_{i, k}\right) =  p\left(\boldsymbol{y}_i\mid \boldsymbol\gamma_{i, k}, \boldsymbol{\chi} \right) = \int p \left(\boldsymbol{y}_i\mid \boldsymbol\gamma_{i, k}, \boldsymbol{\beta}_{i, k}, \sigma_{i, k}^2,\boldsymbol{\chi} \right)\, p\left(\boldsymbol{\beta}_{i, k}, \sigma_{i, k}^2 \mid \boldsymbol\gamma_{i, k},\boldsymbol{\chi} \right) \, d\boldsymbol{\beta}_{i, k}\, d\sigma_{i, k}^2
\end{align*} 
is the marginal likelihood of $\boldsymbol\gamma_{i, k}$.

The most accurate approximation arises when $\boldsymbol\gamma_{i, 1}, \boldsymbol\gamma_{i, 2}, \dots, \boldsymbol\gamma_{i, K}$ are the $K$ highest posterior probability models and the approximation will be accurate if these $K$ models include most of the posterior model probability. 
Although, we cannot guarantee finding these $K$ highest probability models, we will discuss an algorithm in Section~\ref{novel2} for finding models with higher posterior probabilities. 
Substituting the Occam's window approximation in \eqref{approx1} 
into $Q(\boldsymbol\chi)$ in \eqref{simple_EM}
gives 
\begin{align}\label{analysed Q}
Q(\boldsymbol\chi) =& \sum_{i=1}^M 
\E_{\boldsymbol\gamma_i\mid \boldsymbol{\chi}, \boldsymbol{y}_i}\left[
\E_{\boldsymbol\beta_{i}, \sigma_{i}^2 \mid \boldsymbol\gamma_{i}, \boldsymbol{\chi}, \boldsymbol{y}_i}
\left[ \log P\left(\boldsymbol{y}_i\mid \boldsymbol\gamma_{i}, 
    \boldsymbol\beta_{i}, \sigma_{i}^2,\boldsymbol{\chi} \right)+ \log P(\boldsymbol\gamma_{i}, \,
    \boldsymbol\beta_{i}, \sigma_{i}^2\mid \boldsymbol\chi)  \right]\right]\notag\\
&    + \log P(\boldsymbol{\chi})\notag\\
=& \sum_{i=1}^M 
\sum_{k=1}^K w_{i,k} \,
\E_{\boldsymbol\beta_{i, k}, \sigma_{i, k}^2 \mid \boldsymbol\gamma_{i, k}, \boldsymbol{\chi}, \boldsymbol{y}_i}
\left[ \log P\left(\boldsymbol{y}_i\mid \boldsymbol\gamma_{i, k}, 
    \boldsymbol\beta_{i, k}, \sigma_{i, k}^2,\boldsymbol{\chi} \right)+ \log P(\boldsymbol\gamma_{i, k}, \,
    \boldsymbol\beta_{i, k}, \sigma_{i, k}^2\mid \boldsymbol\chi)  \right]\notag\\
    &+ \log P(\boldsymbol{\chi}).
\end{align}

To evaluate these expectations,
we define residuals  $\boldsymbol{r}_{i} = \boldsymbol{y}_{i}- \boldsymbol{X}_{i}\,\boldsymbol{\zeta}^*$ and, for the $k$-th model, define
 $\boldsymbol{A}_{i, k} = \boldsymbol{B}_{i, k} \, (\boldsymbol{S}_{i, k})^T \,\boldsymbol{r}_i$,
 $\boldsymbol{B}_{i, k} = \left((\boldsymbol{S}_{i, k})^T \,\boldsymbol{S}_{i, k} + \boldsymbol{\Lambda}_i\right)^{-1}$, 
 $\boldsymbol{C}_{i, k} =  \boldsymbol{r}_i^{T}\left( I_{n_i} -\boldsymbol{S}_{i, k}
 \,\boldsymbol{B}_{i, k}\,\boldsymbol{S}_{i, k}^{T} \right)\,
 \boldsymbol{r}_i$,
 $m_{i, k} = (\boldsymbol{A}_{i, k})_1$, $\boldsymbol{M}_{i, k} = (\boldsymbol{A}_{i, k})_{2:(p_{i, k}+1)}$,
$q_{i, k} = (\boldsymbol{B}_{i, k})_{1,1}$ and $\boldsymbol{Q}_{i, k} =  (\boldsymbol{B}_{i, k})_{2:(p_{i, k}+1), 2:(p_{i, k}+1)}$ where $\boldsymbol{\Lambda}_{i, k} =  \mbox{diag}(\psi^{-1}, \underbrace{g^{-1}, \dots, g^{-1}}_{p_{i, k}^{\gamma} times})$.\\
The model weight defined in \eqref{weight} is
\begin{align*}
w_{i,k} &\propto   p(\boldsymbol\gamma_{i, k})\,p(\boldsymbol{y}_i\mid \boldsymbol\gamma_{i, k}, \boldsymbol{\chi})\notag \\ 
 &= \Gamma\left(p_{i, k} + a_1\right)\Gamma\left( p-p_{i, k} + b_1 \right)\, (g^2)^{-p_{i, k}^{\gamma}/2} \mid(\boldsymbol{B}_{i, k})^{-1}\mid^{-1/2}\left(b+ \boldsymbol{C}_{i, k}/2\right)^{-(a + n_i)}.
  \label{marg_post}
\end{align*}
The posterior distribution $p(\boldsymbol{\beta}_{i, k}, \sigma_{i, k}^2 \mid \boldsymbol\gamma_{i, k}, \boldsymbol{\chi}, \boldsymbol{y}_i)$ can be factorized as  
$
\sigma_{i, k}^2\mid \boldsymbol\gamma_{i, k}, \boldsymbol{\chi},\boldsymbol{y}_i\sim \mathcal{IG}\left(a + n_i, b + \boldsymbol{C}_{i, k}/2\right)
$ 
and 
$\boldsymbol{\beta}_{i, k}\vert \boldsymbol\gamma_{i, k}, \sigma_{i, k}^2, \boldsymbol{\chi},\boldsymbol{y}_i \sim \N\left(\boldsymbol{A}_{i, k}, \sigma_{i, k}^2 \boldsymbol{B}_{i, k}\right)$. The posterior expectations that we need to evaluate $Q(\boldsymbol\chi)$ are
\begin{align*}
&\E\left[\frac{\boldsymbol{\beta}_{i, k}}{\sigma_{i, k}^2} \right] =
\frac{a + n_i }
{b + \boldsymbol{C}_{i, k}/2}
\,\boldsymbol{A}_{i, k}, \quad
\E\left[
\frac{\boldsymbol{\beta}_{i,k,1}^T\boldsymbol{\beta}_{i,k,1}}{\sigma_{i, k}^2}\right]
 =   \left(q_{i, k} + 
\frac{a + n_i }
{b + \boldsymbol{C}_{i, k}/2}
\left(m_{i, k}\right)^2\right),\\
 &\E\left[
\frac{\boldsymbol{\beta}_{i,k,2:(p_{i, k}+1)}^T\boldsymbol{\beta}_{i,k,2:(p_{i, k}+1)}}{\sigma_{i, k}^2}\right]
=   \left(\mbox{tr}\left(\boldsymbol{Q}_{i, k}\right) + 
\frac{a + n_i }
{b + \boldsymbol{C}_{i, k}/2}
\left(\boldsymbol{M}_{i, k}\right)^T\left(\boldsymbol{M}_{i, k}\right)\right),\\
&\E\left[\frac{1}{\sigma_{i, k}^2}\right] = 
\frac{a + n_i }
{b + \boldsymbol{C}_{i, k}/2},\quad
\E\left[ -  \log(\sigma_{i, k}^2)\right]
= \psi(a + n_i) - \log(b + \boldsymbol{C}_{i, k}/2), \quad 
\E\left[\frac{1}{u}
\right] = \frac{g^2}{1 + g^2}
\end{align*}
where $\psi(\cdot)$ is the digamma function.


The maximizers of $Q(\boldsymbol\chi)$ are available in closed-form for some of the parameters. The maximizer of $\boldsymbol\zeta^{\star}$ and  $\psi$ are
\[
\boldsymbol\zeta^{\star} = \left(
\sum_{i=1}^M  
\boldsymbol{X}_{i}^T \boldsymbol{X}_{i} \sum_{k=1}^K w_{i, k} \, \E\left[ \frac{1}{\sigma_{i, k}^2}\right] \right)^{-1}
\left(\sum_{i=1}^M \left(\boldsymbol{X}_{i}\right)^T \sum_{k=1}^K w_{i, k}\, \left(\E\left[\frac{1}{\sigma_{i, k}^2}\right]\boldsymbol{y}_{i} - \boldsymbol{S}_{i, k} \E\left[\frac{\boldsymbol{\beta}_{i, k}}{\sigma_{i, k}^2} \right] \right)
\right)
\]
and
\[
\psi = \frac{1}{M+4}\left( \sum_{i=1}^M
\sum_{k=1}^K w_{i, k}\,
\E\left[
\frac{\beta_{i,k,1}^2}{\sigma_{i, k}^2}\right] +2\right).
\]

To find the maximizers of $a$ and $b$, we solve the following equations:
\begin{align}\label{popul_prmters2}
\frac{\Gamma'(a)}{\Gamma(a)} =  \log b + \frac{1}{M}\sum_{i=1}^M \sum_{k=1}^K w_{i, k}\,\E\left[ \log\left(\frac{1}{\sigma_{i, k}^2}\right)\right],  \quad b = \frac{a\,M}{\sum_{i=1}^M\sum_{k=1}^K w_{i, k}\, \E\left[ \frac{1}{\sigma_{i,k}^2}\right]}
\end{align}
In the same way, we update to $a_1$ to the maximizer of the equation
\[
 \log\Gamma(a_1+b_1) -\log\Gamma\left(p + a_1 + b_1\right)
- \log\Gamma(a_1)
 +\frac{1}{M}\sum_{i=1}^M \sum_{k=1}^K w_{i, k}\,\log \Gamma\left(p_{i, k} + a_1\right),
\]
 $b_1$ to the maximizer of the equation
\[
 \log\Gamma(a_1+b_1) -\log\Gamma\left(p + a_1 + b_1\right)
- \log\Gamma(b_1)
  +\frac{1}{M}\sum_{i=1}^M \sum_{k=1}^K w_{i, k}\, \log\Gamma\left(p - p_{i, k} + b_1\right)
\]
and $g$ to the maximizer of the equation
\[
 - \log g\sum_{i=1}^M \sum_{k=1}^K w_{i,k}\,  p_{i, k}
  - \frac{1}{g}\sum_{i=1}^M \sum_{k=1}^K w_{i,k}\, \E_{\boldsymbol{\beta}_{i, k}, \sigma^2_{i, k}\mid \boldsymbol\gamma_{i, k},\boldsymbol{\chi}, \boldsymbol{y}_i} \left[  \frac{\sum_{j=1}^{p_{i. k}}\boldsymbol{\beta}_{i, k, j}^2}{\sigma_{i, k}^2}\right] 
 -  \log g - 2\log(1 + g).
\]

The calculation of these values  can be speeded up by only summing  over the $k$ such that $w_{i, k} > \epsilon$ where $\epsilon$ is chosen by the user to be small. This corresponds to using 
\begin{equation*}
Q(\boldsymbol\chi)
= \sum_{i=1}^M 
\sum_{k=1}^K \mbox{I}(w_{i,k} > \epsilon)
w_{i,k} \,
\E_{\boldsymbol\beta_{i, k}, \sigma_{i, k}^2 \mid \boldsymbol\gamma_{i, k}, \boldsymbol{\chi}, \boldsymbol{y}_i}
\left[ \log P\left(\boldsymbol{y}_i\mid \boldsymbol\gamma_{i, k}, 
    \boldsymbol\beta_{i, k}, \sigma_{i, k}^2,\boldsymbol{\chi} \right)+ \log P(\boldsymbol\gamma_{i, k}, \,
    \boldsymbol\beta_{i, k}, \sigma_{i, k}^2\mid \boldsymbol\chi)  \right].
\end{equation*} 
 
\subsection{Initialization} \label{novel1}

The convergence and computational time of the EM algorithm can depend on initialisation of $\boldsymbol\chi$. We use the following scheme to initialize the elements of $\boldsymbol\chi$:
\begin{itemize}
    \item $\zeta^{\star} = \left(\sum_{i=1}^M \boldsymbol{X}^{\star\,T}_i \boldsymbol{X}^{\star}_i\right)^{-1}\left(\sum_{i=1}^M \boldsymbol{X}^{\star\,T}_i \boldsymbol{y}^{\star}_i\right)$ where $\boldsymbol{y}^{\star}_i = \boldsymbol{y}_i - \bar{\boldsymbol{y}}_i$.
    \item To initialize $b$, we estimate the LME model
    \[
    \boldsymbol{r}_i = \boldsymbol{y}_i - \boldsymbol{X}_i\boldsymbol\zeta^{\star} = \mu_i + \boldsymbol{\epsilon}_i
    \]
    where $\boldsymbol{\epsilon}_i \sim \N(0, \sigma^2 I_{n_i})$ and 
    set  $a=$ and $b=a\hat\sigma^2$ where $\hat\sigma^2$ is the estimate of $\sigma^2$.

    \item There are $\left(\begin{array}{c} p\\ k\end{array}\right)$ possible sub-models of $\boldsymbol{S}_i$ with $k$ variables. We find the smallest $p^{\star}$ for which $\sum_{k=1}^{p^{\star}} \left(\begin{array}{c} p\\ k\end{array}\right)\geq K$. For $i=1,\dots, M$, we calculate the marginal likelihood for model with $1,\dots, p^{\star}$ possible variables and initialize Occam's window for the $i$-th individual with $K$ model with the highest marginal likelihoods.

    
    \end{itemize}

\subsection{Updating Occam's window} \label{novel2}

The success of the algorithm depends on finding high probability models in Occam's window and a greedy search algorithm is used to achieve this. We define $\tilde{m}_i = \min\{m(\boldsymbol\gamma_{i, k})\}$ and  update Occam's window for the $i$-th individual using the following steps:
\begin{enumerate}
\item choose a model uniformly at random. Let that model be $\boldsymbol\gamma_{i, k}$.
\item choose a variable $j$, uniformly at random from $1, \dots, p$.
\item propose a new model $\boldsymbol{\tilde\gamma}$  with the values $\boldsymbol{\tilde\gamma}_{j} = 1 - \boldsymbol\gamma_{i, k, j}$ and 
$\boldsymbol{\tilde\gamma}_{m} = \boldsymbol\gamma_{i, k, m}$ for $m \neq j$. The new model $\boldsymbol{\tilde\gamma}$ will include variable $j$ if it is excluded from $\boldsymbol{\tilde\gamma}_{i, k, j}$ or vice versa.
\item check that $\boldsymbol{\tilde\gamma}$ is not in Occam's window. If it is not, include  $\boldsymbol{\tilde\gamma}$ in Occam's window if 
$m_i(\boldsymbol{\tilde\gamma}) > \tilde{m}_i$ by replacing the model corresponding to the value $ \tilde{m}_i$ and 
then re-calculate $\tilde{m}_i$.
\end{enumerate}

We could choose to use same number of update of Occam's window 
for each athlete in each iteration of the algorithm. However, we choose to improve efficiency by only updating $L$ models across all athletes in one iteration of the algorithm and biasing the number of updates towards individuals where changes are more likely to be accepted. Define $t_i$ to be the number of updates since the previous successful update and at the start $t_i=0$ for all athletes. We sample $r_i \sim \mbox{Ex}(1 + t_i)$ and 
$p_i = \frac{r_i}{\sum_{j=1}^M r_j}$
for $i=1,\dots, M$.
The probabilities $p_i$ will tend to be larger for individuals who are regularly updated as for those $t_i$ will be set to zero and hence
 $\E[r_i] = \frac{1}{1 + t_i}$.  Define $l_i$ to be the number of times that Occam's window for the $i$-th individuals is updated and generate
$l_1,\dots, l_M$ from a multiomial distribution with total sample size $L$ and probabilities $p_1, \dots, p_m$.

The algorithm alternates between updating Occam's window and the population-level parameters $\boldsymbol\chi$.



\section{Extending the algorithm to more complicated models}\label{t-skew}

The algorithm developed in the previous section works for inference in the LME model in \eqref{lme} with normal errors but the method can be extended to other error distributions.  As motivation, \cite{griffin2022bayesian} suggested using skew $t$ distributed errors \citep{Azzalini2003} in a linear mixed model for elite athletic performances due to the occurrence of unusually poor  performances. Our Occam's window approach can be extended using a variational Bayes algorithm \citep[see][for a review]{BleiKucMcA17}. Variational Bayes algorithms have been developed for skew-normal and $t$ distributions using latent variable representations. \cite{Wand11} consider inference about the parameters of the skew-normal and $t$ distributions and \cite{Guha15} consider inference in inverse problems with skew-$t$ errors.

We denote the skewness and 
 degrees of freedom parameters by $c$ and $f$. A convenient latent variable representation writes the model in   \eqref{lme} with skew $t$ errors
 in the following form:
\begin{align}\label{skew-t model}
\boldsymbol{y}_{i} = \boldsymbol{X}_{i}\,\boldsymbol{\zeta}^*  + \boldsymbol{S}^{\gamma}_{i}\,\boldsymbol{\beta}^{\gamma}_i 
+ \frac{c}{\sqrt{1+c^2}}\,\boldsymbol{d}_{i} +\boldsymbol{\epsilon}^*_{i}
\end{align}
where  
\[
\boldsymbol\epsilon_{i}^* \stackrel{ind.}{\sim}\N\left(0, \frac{\sigma^2_i}{\boldsymbol{\rho}_{i}}\frac{1}{1+c^2}\right),\quad 
\boldsymbol{d}_{i}\stackrel{ind.}{\sim} \mathcal{TN}_{[0,\infty]} \left(0, \frac{\sigma^2_i}{\boldsymbol{\rho}_{i}} \right),\quad 
\boldsymbol{\rho}_{i} \sim 
\mathcal{G}a\left(\frac{f}{2},\frac{f}{2}\right)\]
and division by a vector refers to element-wise division. The introduction of latent variables
$\boldsymbol{\rho}_i$ and $\boldsymbol{d}_i$ where 
$\boldsymbol{\rho}_i = (\rho_{i,1}, \dots, \rho_{i, n_i})$ and  $\boldsymbol{d}_i = (d_{i,1},\dots, d_{i, n_i})$ respectively leads to  a conditionally normal linear model. We will use the prior distributions $f \sim \mathcal{G}a(2,0.1)$ and $c \sim \N(0,10^2)$.

Again, we drop the $\gamma$ superscript notation. The individual-level parameters are now 
$\boldsymbol\nu_i = \{\boldsymbol\gamma_i, \boldsymbol\beta_i, \sigma^2_i, \boldsymbol\rho_i, \boldsymbol{d}_i\}$. The  Occam's window approach developed in Section \ref{infer} can be used in a variational Bayes method for the individual-level parameters. The variational distribution is
\[
q(\boldsymbol\gamma_i, \boldsymbol{\beta}_i,\sigma^2_i,\boldsymbol{\rho}_i, \boldsymbol{d}_i\mid \boldsymbol{\chi}, \boldsymbol{y}_i) =  
q_{\boldsymbol\psi_i}(\boldsymbol{\rho}_i, \boldsymbol{d}_i)
\,q_{\boldsymbol{w}_i,\boldsymbol\phi_i}(\boldsymbol\gamma_i, \boldsymbol{\beta}_i,\sigma^2_i)
\]
where $\boldsymbol\psi_i$, $\boldsymbol{w}_i$ and $\boldsymbol\phi_i$ are variational parameters.
We write $\E_{\boldsymbol\psi_i}$ and $\E_{\boldsymbol\phi_i}$ as expectations with respect to $q_{\boldsymbol\psi_i}$ and $q_{\boldsymbol\phi_i}$ respectively.
The mean-field variational distribution for $q_{\phi}(\boldsymbol\gamma_i, \boldsymbol{\beta}_i,\sigma^2_i)$ 
is proportional to 
\begin{align*}
&- \frac{1}{2\sigma_i^2} \sum_{j=1}^{n_i} \E_{\boldsymbol\psi_i}[\rho_{i, j}] \left[( \sqrt{1+c^2}\left(y_{i, j} - \boldsymbol{X}_{i, j}\boldsymbol{\zeta}^*\right)-
\sqrt{1+c^2}\boldsymbol{S}_{i, j}^{\gamma}\boldsymbol{\beta}_i^{\gamma}
 - \frac{1}{\E_{\boldsymbol\psi_i}[\rho_{i, j}]}c  \E_{\boldsymbol\psi_i}\left[ \rho_{i, j} d_{i, j}\right]\right]^2\\
&- \frac{1}{2\sigma_i^2} c^2 \sum_{j=1}^{n_i}\E_{\boldsymbol\psi_i}\left[
 \rho_{i, j} d_{i, j}^2\right]
+ \frac{1}{2\sigma_i^2} c^2\sum_{j=1}^{n_i} \frac{\E_{\boldsymbol\psi_i}\left[ \rho_{i, j} d_{i, j}\right]^2}{\E_{\boldsymbol\psi_i}[\rho_{i, j}]}
 - \frac{1}{2\sigma^2_i}\sum_{j=1}^{n_i} \E_{\boldsymbol\psi_i}\left[\rho_{i, j} d_{i,j}^{2}\right]\\
 & -  n_i  \log \sigma_i^2 
 -  \frac{1}{2} \log(\psi\sigma_i^2) -\frac{1}{2}\frac{(\boldsymbol{\beta}_{i}^{\gamma})^T\boldsymbol{\Lambda}_i\boldsymbol{\beta}_i^{\gamma}}{\sigma_i^2}
- \frac{p^{\gamma}_i}{2}\log (g^2\sigma_i^2)
\end{align*}
and we use the Occam's window approach to approximate this distribution.
We use the mean-field approximation of the distribution $q_{\psi_i}(\boldsymbol{\rho}_i, \boldsymbol{d}_i)$.
Unlike \cite{Guha15} who factorize this distribution 
as $q_{\boldsymbol{\rho}_i}(\boldsymbol{\rho}_i)\,
q_{\boldsymbol{d}_i}(\boldsymbol{d}_i)$, we derive their joint mean-field distribution which can be factorized as 
$\prod_{j=1}^{n_i} q_{\psi_{i, j}}(\rho_{i, j}\mid d_{i, j})
q_{\psi_{i, j}}(d_{i, j})$. The density $q_{\psi_{i, j}}(d_{i, j})$ is proportional to
\[
\I(d_{i, j} > 0)
\left(
\frac{\frac{f+1}{\nu} \left (d_{i, j} -  \mu
\right)^2}{f+1} +1\right)^{-(\frac{f+1}{2} +\frac{1}{2})}
\]
where
\[
\mu = \frac{   c\sqrt{1+c^2}  
\sum_{k=1}^K  w_{i,k}\,
 \E_{\boldsymbol\phi_i}\left[\frac{1}{\sigma_i^2}
 \left(y_{i, j} - \boldsymbol{X}_{i, j}\boldsymbol{\zeta}^*- \boldsymbol{S}_{i, j}^{\gamma}\boldsymbol{\beta}_i^{\gamma}\right) \right]}
 {(1+c^2)\sum_{k=1}^K  w_{i,k}\,
  \E_{\boldsymbol\phi_i}\left[\frac{1}{\sigma_i^2}
\right]},
\]
\[
\frac{1}{\nu} = \frac{1+c^2}{\lambda} \sum_{k=1}^K  w_{i,k} \,
\E_{\boldsymbol\phi_i}\left[\frac{1}{\sigma_i^2}\right],
\]
which is a truncated $t$-distribution. The distribution
$q_{\psi_{i, j}}(\rho_{i, j}\mid d_{i, j}) = \mbox{Gamma}\left(a_{i, j}^{\star}, b_{i, j}^{\star}\right)$ where 
$a = \frac{1+f}{2}$ and 
\[
b =  \frac{1}{2}\sum_{k=1}^K w_{i,k} \, \E_{\boldsymbol\phi_i}\left[\frac{1}{\sigma_i^2} 
\left( \sqrt{1+c^2}\left(y_{i, j} - \boldsymbol{X}_{i, j}\boldsymbol{\zeta}^*- \boldsymbol{S}_{i}^{\gamma}\boldsymbol{\beta}_i^{\gamma}\right)- c d_{i, j}\right)^2 + \frac{f}{2}\right].
\]

The global parameter $\boldsymbol\chi$ are estimated by maximising the 
$Q$ function 
\begin{equation}\label{analysed Q2}
Q(\boldsymbol{\chi}) =  \sum_{i=1}^M  \sum_{k=1}^K w_{i, k}\E_{\boldsymbol\psi_i}
\left[
\E_{\boldsymbol\phi_i} \left[ \log P\left(\boldsymbol{y}_i\mid \boldsymbol{\nu}_i,\boldsymbol{\chi} \right)+ \log P(\boldsymbol{\nu}_i \mid \boldsymbol\chi)\right] \right] +\log P(\boldsymbol{\chi}).
\end{equation}
Expectations of function of $\boldsymbol\rho_i$ and $\boldsymbol{d}_i$ are approximated using Monte Carlo averages.

The algorithm updates parameters in three blocks
\begin{enumerate}
\item Update $\boldsymbol\chi$ by finding the values that maximize $Q(\chi)$ in 
\eqref{analysed Q2}.

\item Simulate $\boldsymbol{\rho}_i$ and $\boldsymbol{d}_i$ for $i = 1,\dots, M$ and calculate Monte Carlo average of functions of $\boldsymbol{\rho}_i$ and $\boldsymbol{d}_i$ needed to evaluate model probabilities.
\item Update Occam's window.
\end{enumerate}

\section{Examples and Illustrations}\label{results}
\subsection{Simulation Study}

To understand the performance of the algorithms. We performed a simulation study using both the model with skew t errors and different values of model parameters. In each data set, there were 300 individuals, $\zeta_0 = 0$ and $\zeta_i \sim \N(0, 1)$ for $i=1,\dots, 5$. For the $i$-th individual we choose between a large number of observations $n_i = 200$ with probability $q$ and a small number of observations $n_i = 50$ otherwise. The regressors are independent with 
$X_{i, j}\sim \mathcal{N}(0, 1)$ and $S_{i, j}\sim \mathcal{N}(0, 1)$. The error variance is generated
$\sigma^2_i\sim \mathcal{IG}(10, 0.1)$
and the coefficients of the random effects are independent with $\beta_{i, k}\sim h \mathcal{N}(0, 1) + (1 - h) \delta_{\beta_{i, k} = 0}$. 



The data sets are formed by different combinations of the following parameters. We consider settings which have small ($p=10$) or large ($p=20$) number of random effects with the sparser ($h=0.1$) or denser ($h=0.25$) coefficients. The proportion of larger data sets for individuals is either $q=0.15$ or $q=0.3$. 
We consider a symmetric version ($c=0$) or fairly heavily skewed ($c=4$) and heavy-tailed ($f=5$) and close to normal tails ($f=20$). This leads to 5 different parameters with 2 possible values leading to 32 combinations.
All results are calculated using 30 replicate data sets and with two possible values of Occam's window $K=30$ and $K=100$.

\begin{figure}[h!]
\begin{center}
Skew t\\
\begin{tabular}{cc}
$f = 5$, $c = 0$ &   $f = 5$, $c = 4$\\
\includegraphics[scale=0.4]{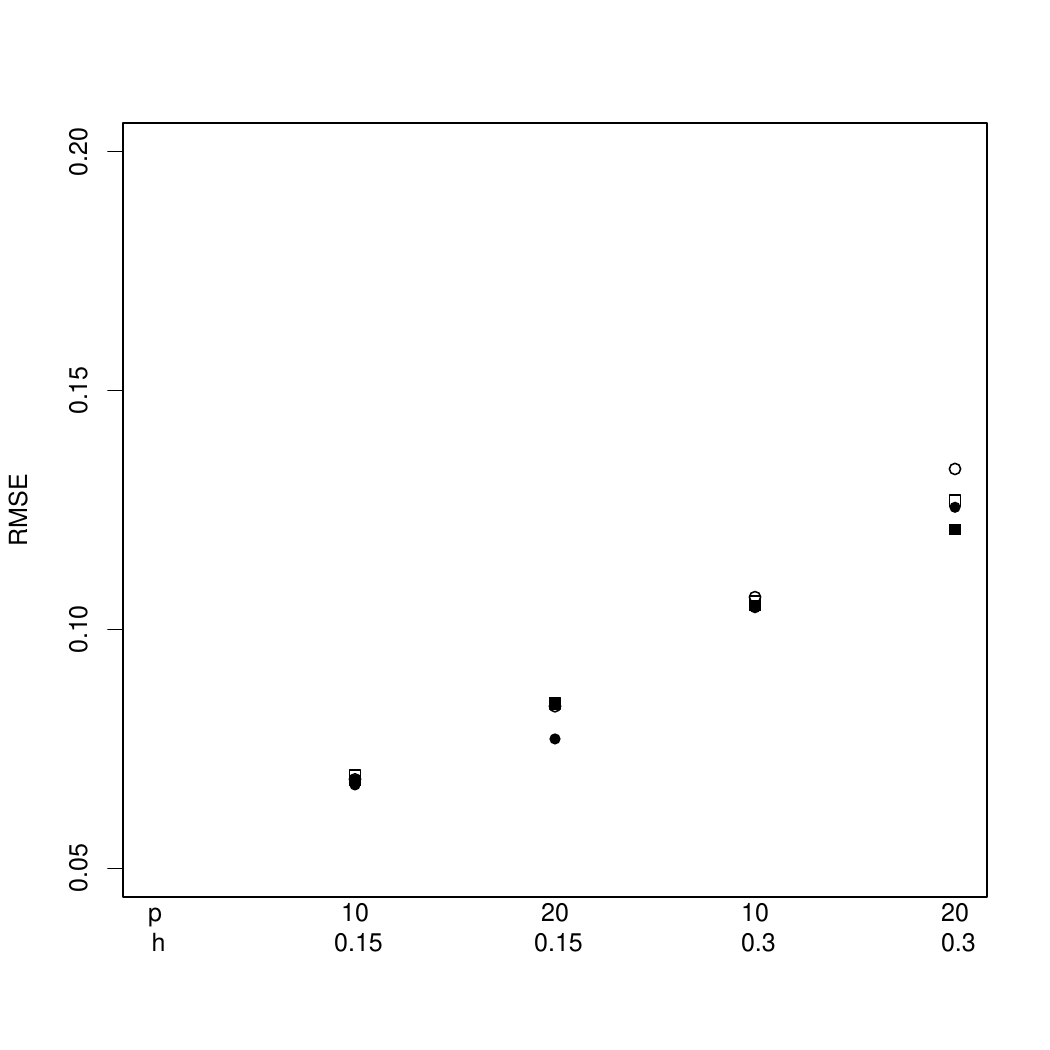} &
\includegraphics[scale=0.4]{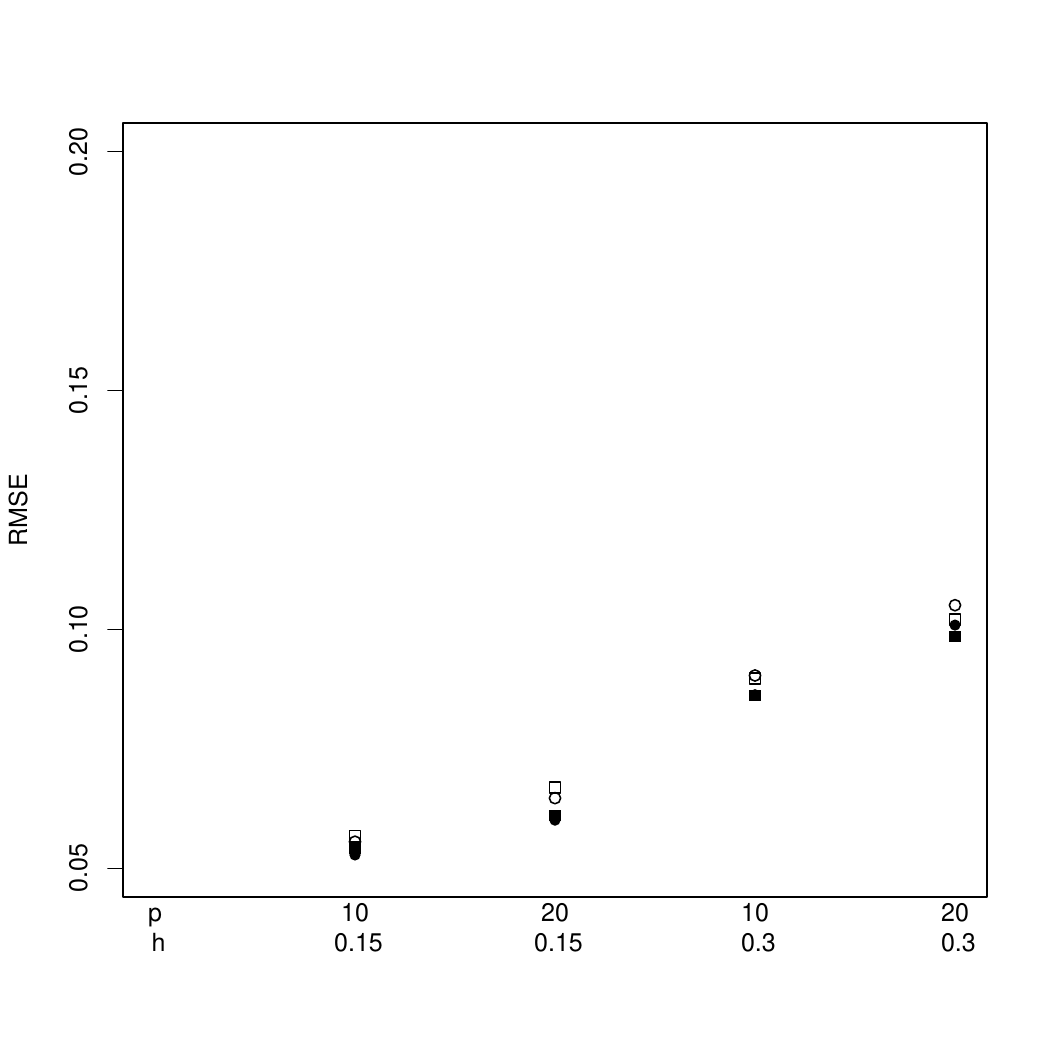}\\
$f = 20$, $c = 0$ &   $f = 20$, $c = 4$\\
\includegraphics[scale=0.4]{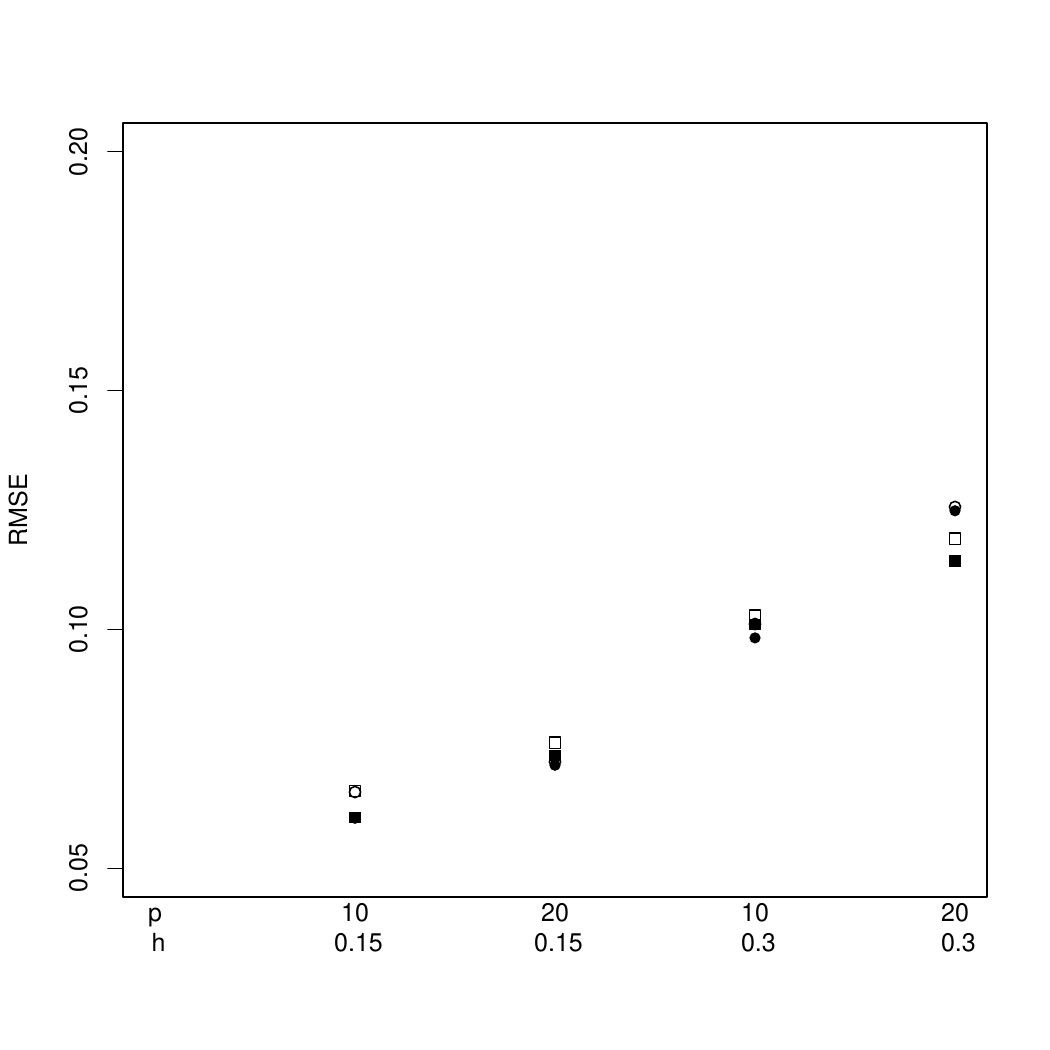} &
\includegraphics[scale=0.4]{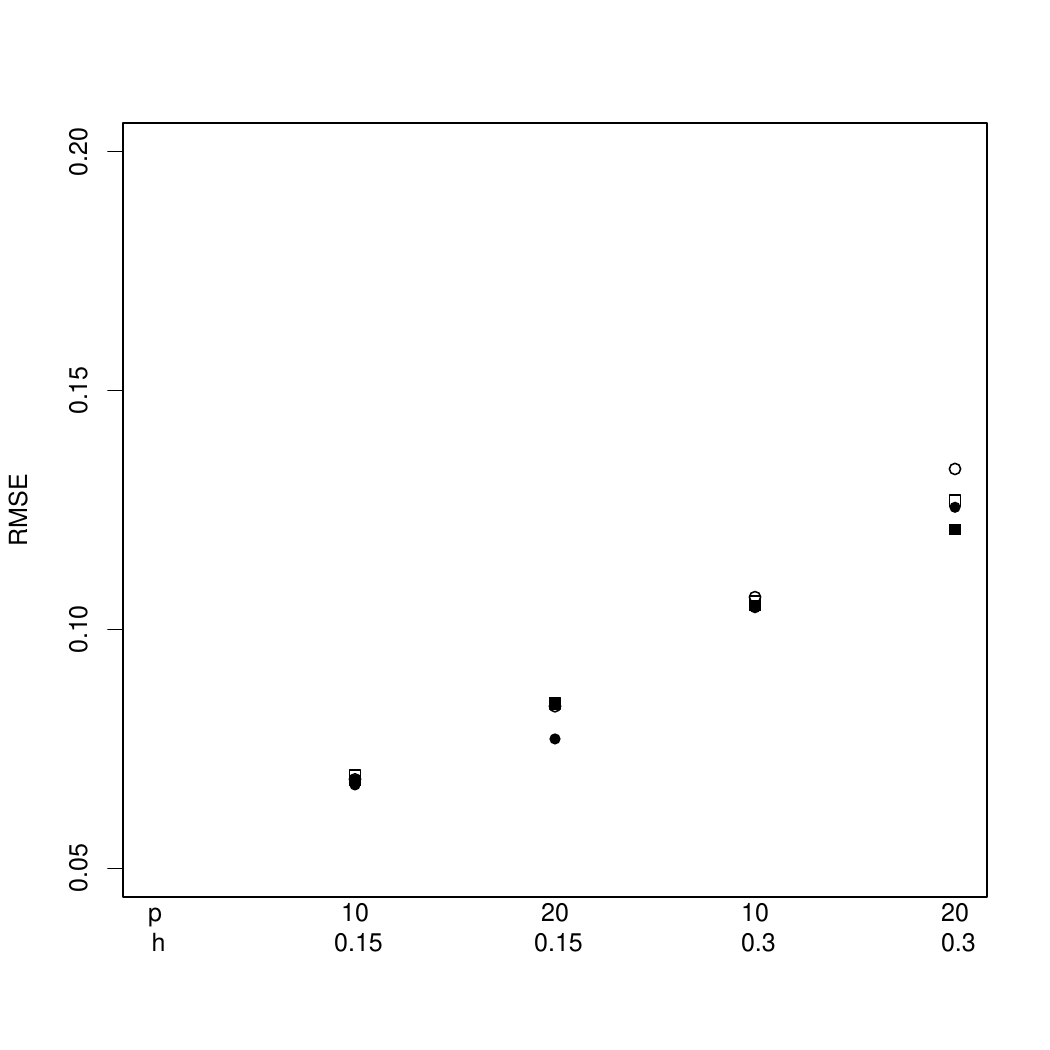}
\end{tabular}
\end{center}
\caption{Simulation study: RMSE for $\gamma$. The symbols represent
 $q = 0.15$ and window = 30 (\raisebox{0.02cm}{\scalebox{0.6}{$\square$}}), 
 $q = 0.15$ and window = 100 ($\circ$),  
 $q = 0.3$ and window = 30 (\raisebox{0.02cm}{\scalebox{0.6}{$\blacksquare$}}), and 
 $q = 0.3$ and window = 100 ($\bullet$)
 }
\end{figure}

We compare the performance using the root mean squared error (RMSE). The RMSEs of $c$ and $f$ are calculated in the usual way. The RMSE of $\gamma$ is computed using
\[
\mbox{RMSE} = \frac{1}{30\,p} \sum_{i=1}^{30}\sum_{k=1}^p (\gamma_k^{(i)} - \E[\gamma_k\mid {\bf y}^{(i)}])^2
\]
where $\gamma_1^{(i)}, \dots, \gamma_p^{(i)}$ and ${\bf y}^{(i)}$ represent the $\gamma$ and the data in the $i$-th replicate. 

The RMSEs of the $\gamma$'s show a similar pattern with all choices of skewness and degrees of freedom. The performance is better with sparser regression coefficients and a smaller number of variables $p$. Performance also tends to be better with a larger proportion of large individual sample sizes ($q=0.3$) but the effect is small. The effect of windows is also small and does not have a consistent effect across the different choices for the error distribution. The RMSE tends to be slightly smaller for skew $t$ distribution with a large skewness ($c=4$0) and degrees of freedom ($f=5$) and slightly larger for the normal error compared to other settings. In the model with skew $t$ errors, the RMSE for the skewness and degrees of freedom are given in the appendix. The skewness is well-estimated for both $c=0$ and $c=4$. The degrees of freedom is unsurprisingly harder to estimate but we show good performance when $f=5$. With $f=20$, the errors are larger but this is partly due to the difficulty of distinguishing distribution with 10 or 20 degrees of freedom.

\begin{figure}[h!]
\begin{center}
Skew t\\
\begin{tabular}{cc}
$f = 5$, $c = 0$ &   $f = 5$, $c = 4$\\
\includegraphics[scale=0.4]{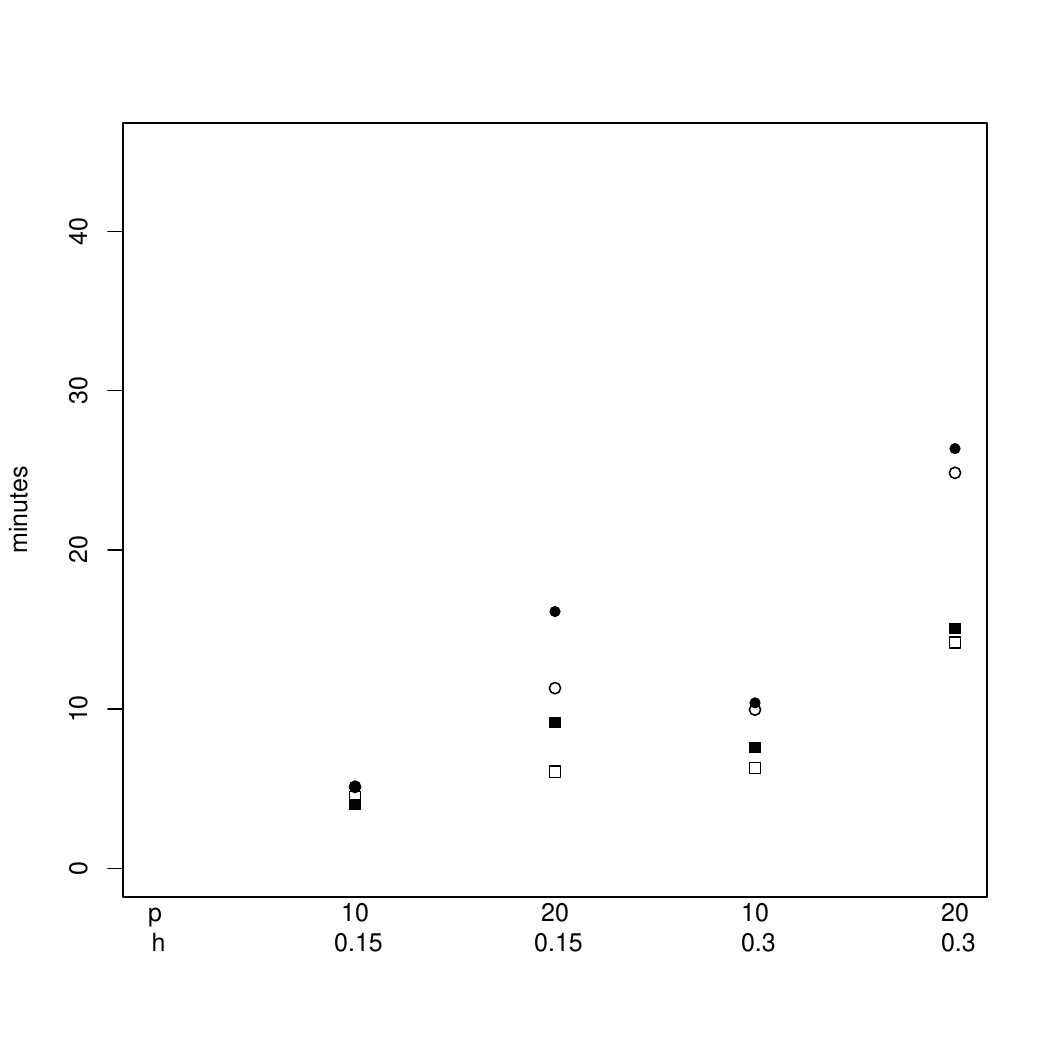} &
\includegraphics[scale=0.4]{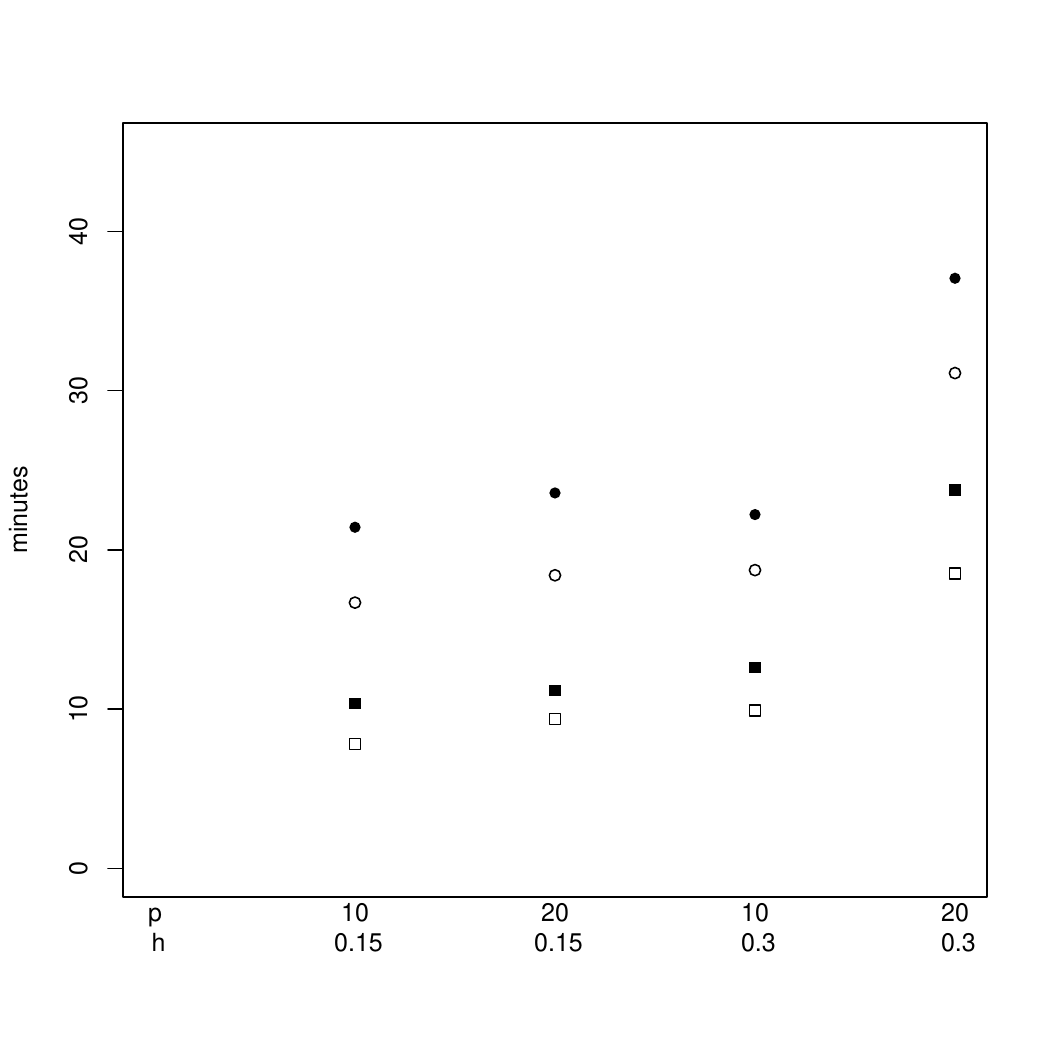}\\
$f = 20$, $c = 0$ &   $f = 20$, $c = 4$\\
\includegraphics[scale=0.4]{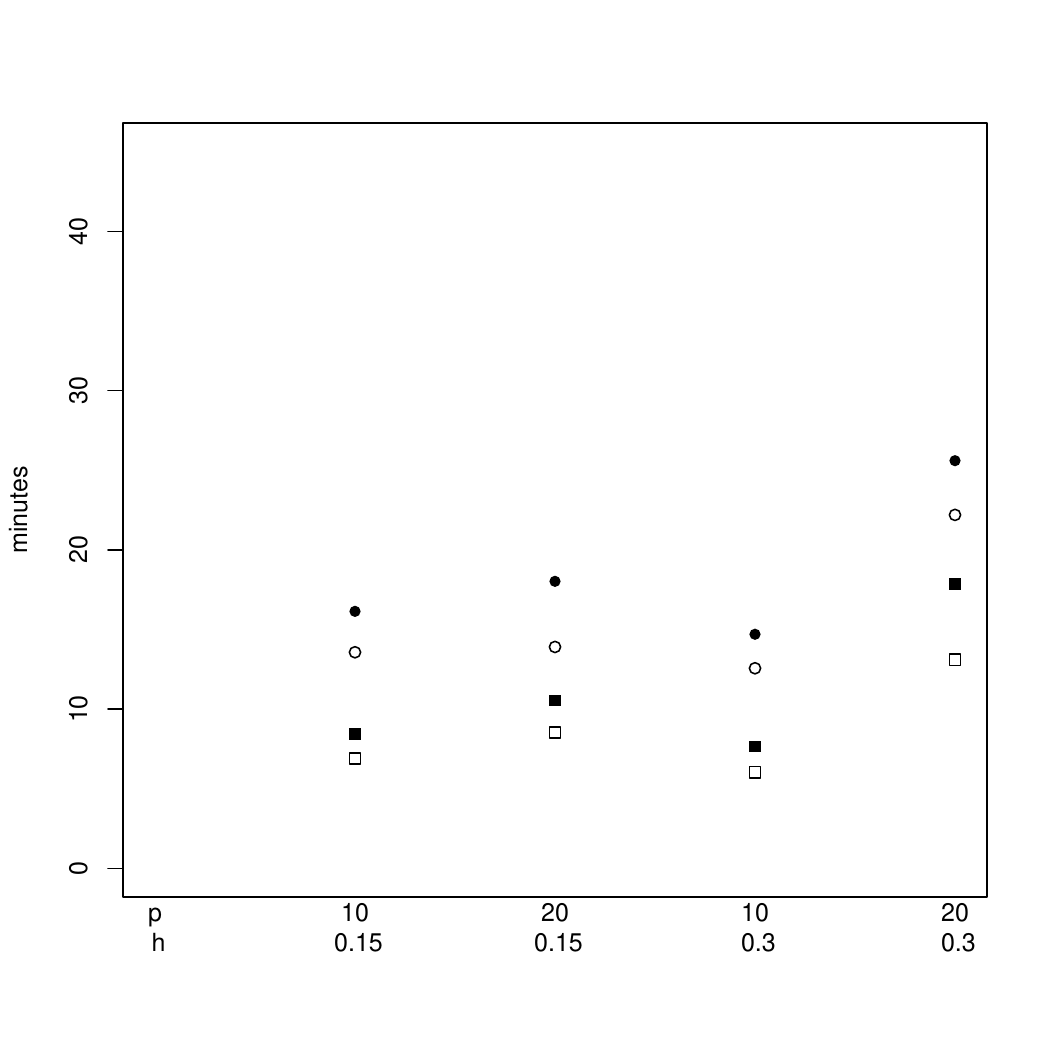} &
\includegraphics[scale=0.4]{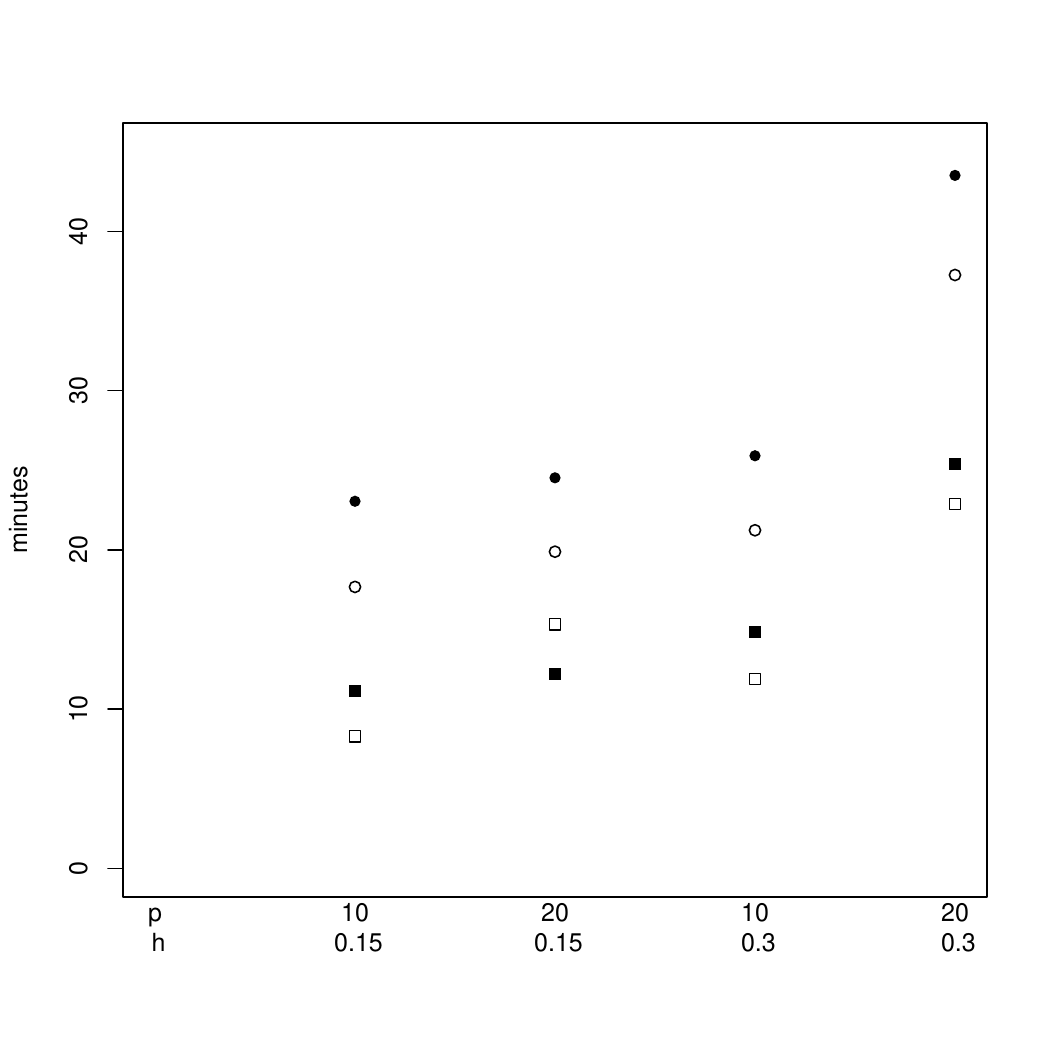}
\end{tabular}
\end{center}
\caption{Simulation study: average time in minute. The symbols represent
 $q = 0.15$ and window = 30 (\raisebox{0.02cm}{\scalebox{0.6}{$\square$}}), 
 $q = 0.15$ and window = 100 ($\circ$),  
 $q = 0.3$ and window = 30 (\raisebox{0.02cm}{\scalebox{0.6}{$\blacksquare$}}), and 
 $q = 0.3$ and window = 100 ($\bullet$)
 }
\end{figure}

\subsection{Results and Application}
\begin{table}[!htbp]
		\caption{EM and MCMC estimates of common-level parameters}
		\label{tab:tab1}
 \begin{center}
 \begin{tabular}{r r r r r r r r}
		\hline
             &&\textbf{EM}&&&\textbf{MCMC}&\\
	&	&Time(min)& Skewness & Dof&Time &Skewness & Dof \\
	   \hline
      100m &   Males &12&0.7&19.05& &1.21&18.4\\
        & Females &35&0.9&20.4&&1.35 &19.6\\
	   \hline
   Weightlifting  &      Males &50&-0.6&17&&-1.91&7.8\\
     &   Female &40&-1.1&5.8&&-1.64 &6.9\\
	        \hline
	   \end{tabular}
	\end{center}
\end{table}

\begin{figure}[h!]
\centering
\begin{tabular}{c c c}
&Males & Females \\
\raisebox{2cm}{100m} &
\includegraphics[scale=0.3]{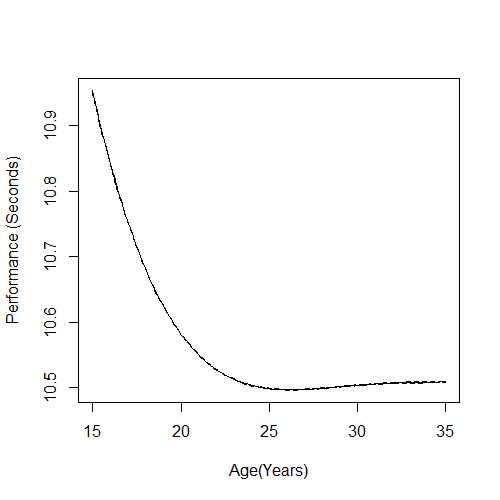}&
\includegraphics[scale=0.3]{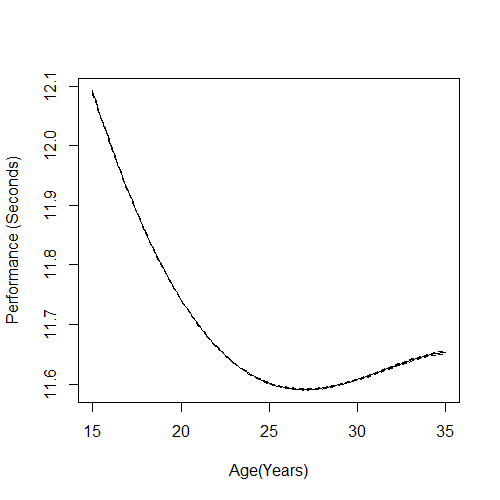}\\
\raisebox{2cm}{Weightlifting} & \includegraphics[scale=0.3]{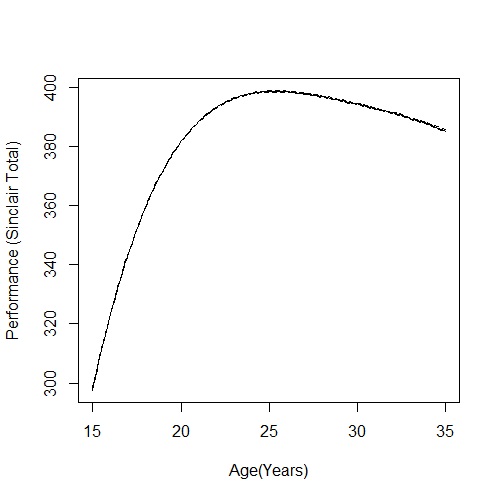}&
\includegraphics[scale=0.3]{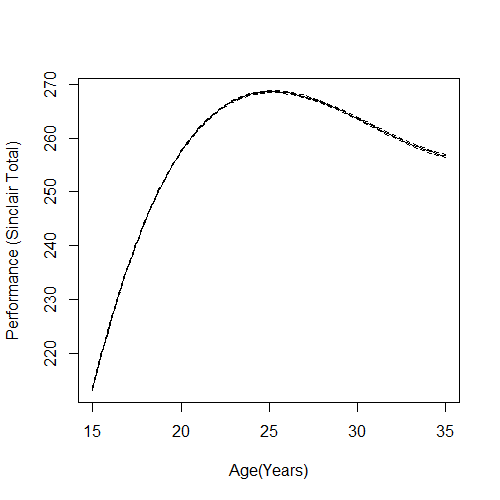}
\end{tabular}
\caption{\label{plt1b} Population performance plot for both Males and Females in 100m and weightlifting}
\end{figure}

\begin{figure}[!htbp]
\centering
\begin{tabular}{c c} 
Males & Females \\
\includegraphics[scale=0.3]{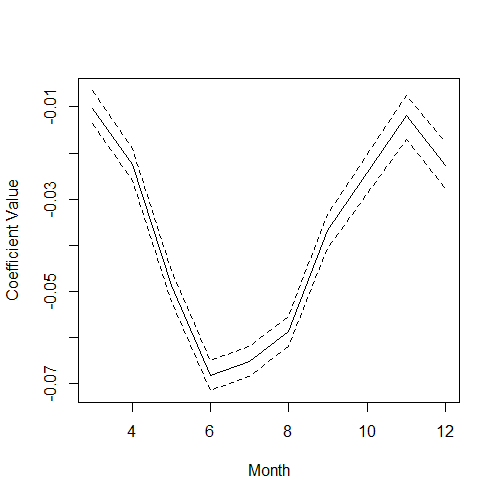}&   \includegraphics[scale=0.3]{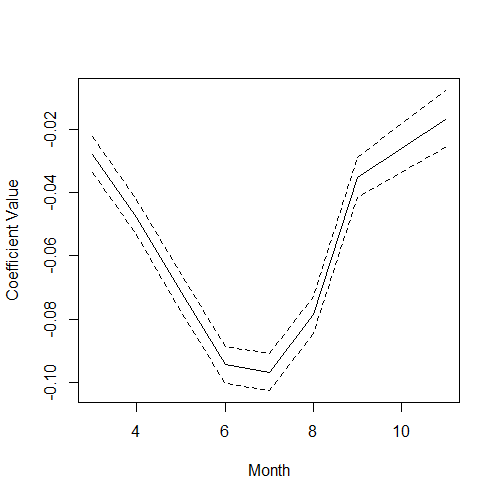}\\
\includegraphics[scale=0.3]{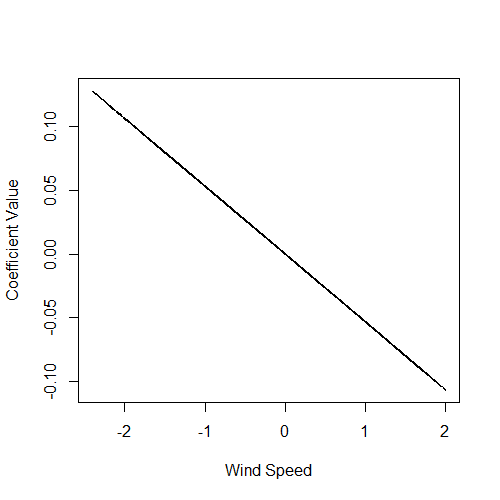}&
\includegraphics[scale =0.3]{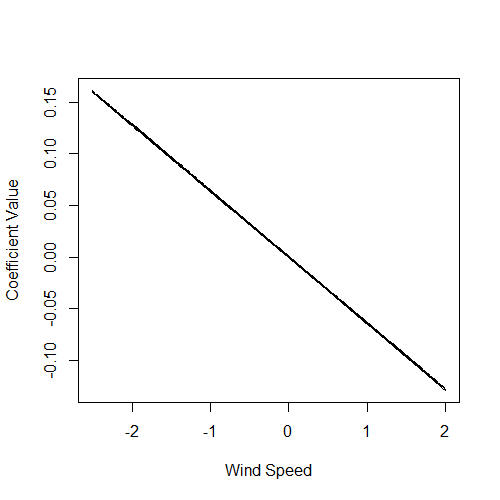}
\end{tabular}
\caption{\label{plt2a} Month and Wind Effect for 100m }
\end{figure}

\begin{figure}[!htbp]
\centering
\begin{tabular}{c c c}
$\text{Athlete 1}$&$\text{Athlete 2}$&$\text{Athlete 3}$\\
\includegraphics[scale=0.3]{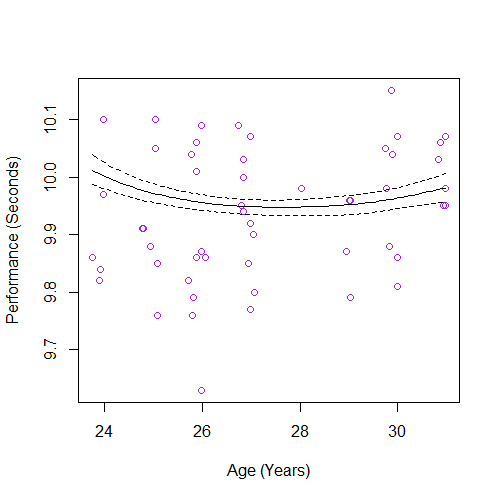}&
\includegraphics[scale=0.3]{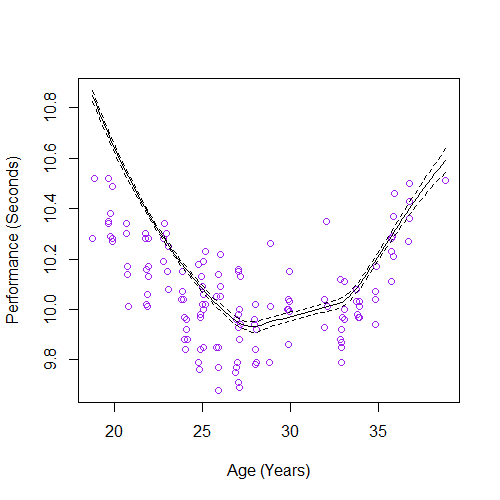}& 
\includegraphics[scale=0.3]{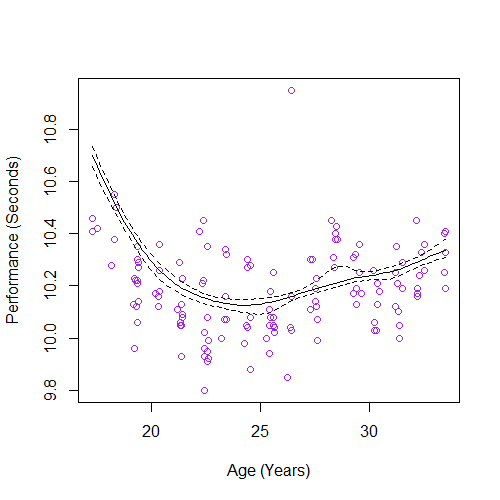}\\
\includegraphics[scale=0.3]{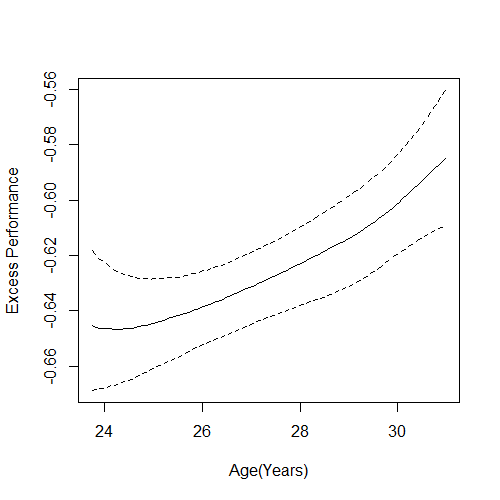}& \includegraphics[scale=0.3]{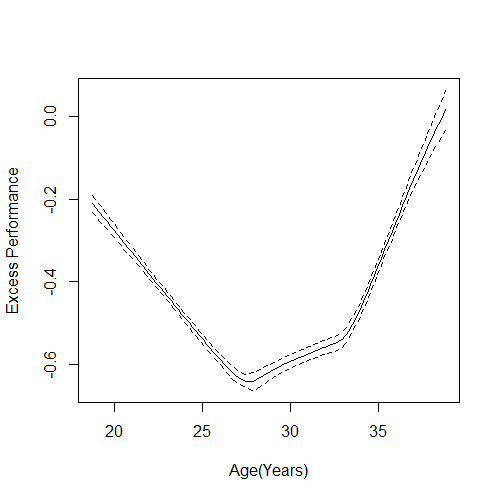}&
\includegraphics[scale=0.3]{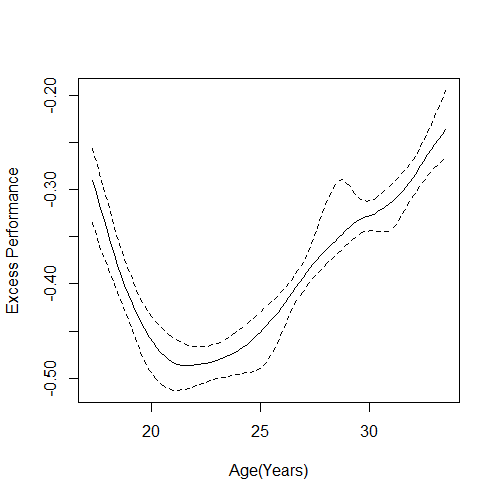}
\end{tabular}
\caption{\label{plt1c} Individual and excess  performance for 100m Male Athletes}
\end{figure}

\begin{figure}[!htbp]
\centering
\begin{tabular}{c c c}
$\text{Athlete 4}$&$\text{Athlete 5}$&$\text{Athlete 6}$\\
\includegraphics[scale=0.3]
{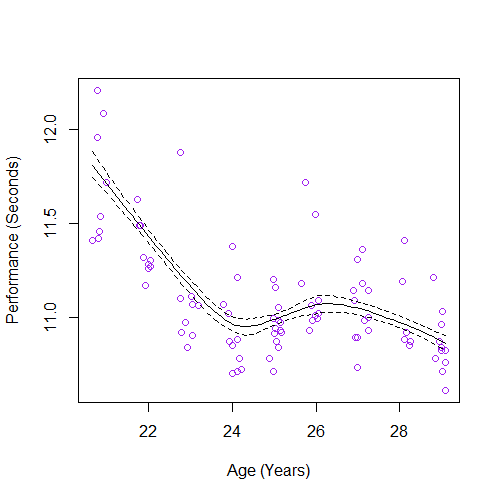}&
\includegraphics[scale=0.3]
{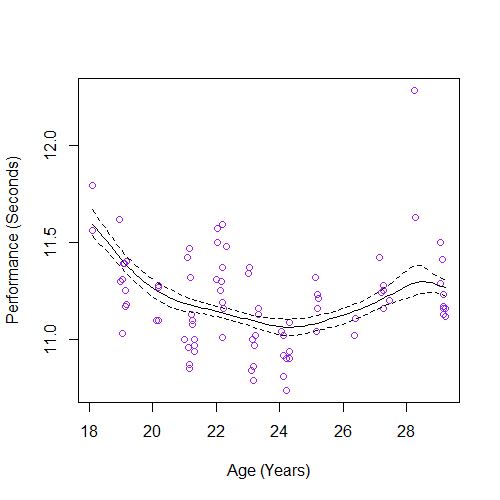}&	
\includegraphics[scale=0.3]{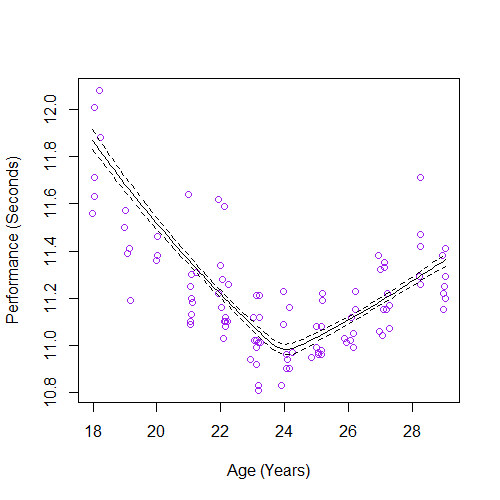}\\
\includegraphics[scale=0.3]{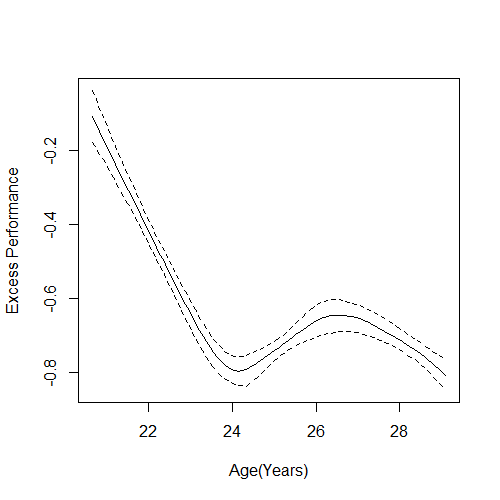}&
\includegraphics[scale=0.3]{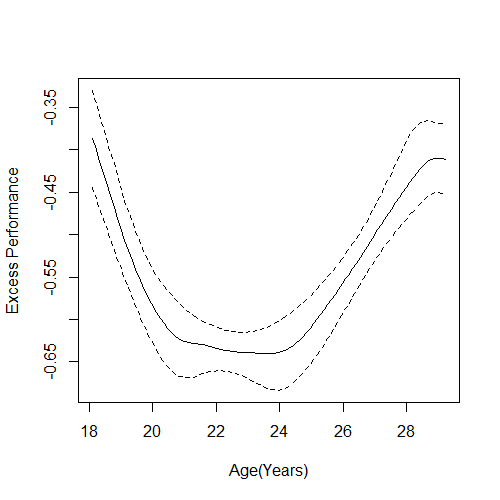}&	
\includegraphics[scale=0.3]{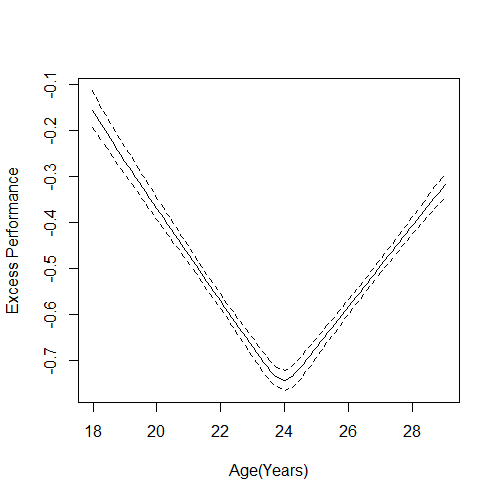}
\end{tabular}
\caption{\label{plt1d} Individual and excess  performance for 100m Female Athletes}
\end{figure}

\begin{figure}[!htbp]
\centering
\begin{tabular}{c c c}
$\text{Athlete 7}$&$\text{Athlete 8}$&$\text{Athlete 9}$\\
\includegraphics[scale=0.3]{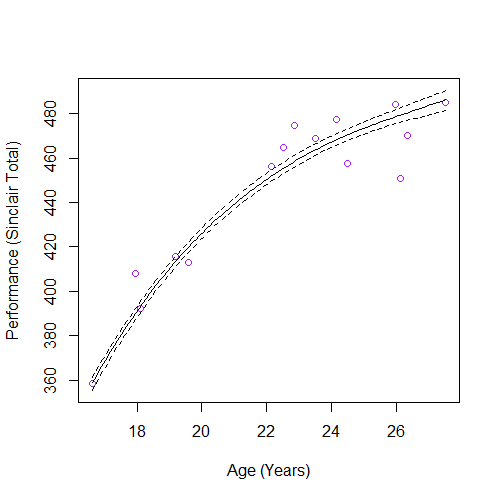}&
\includegraphics[scale=0.3]{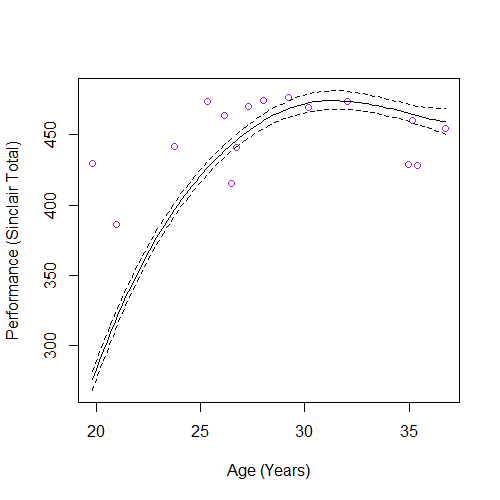}&
\includegraphics[scale=0.3] {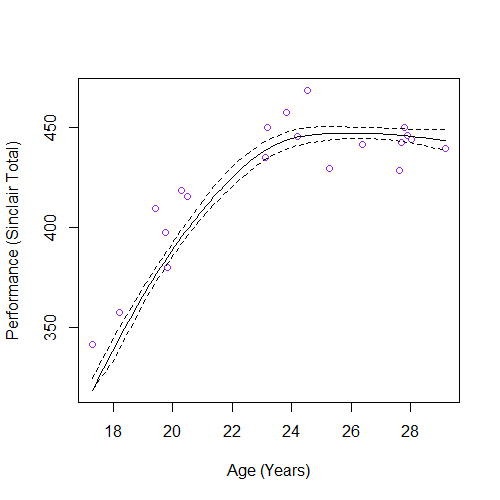}\\	
\includegraphics[scale=0.3]{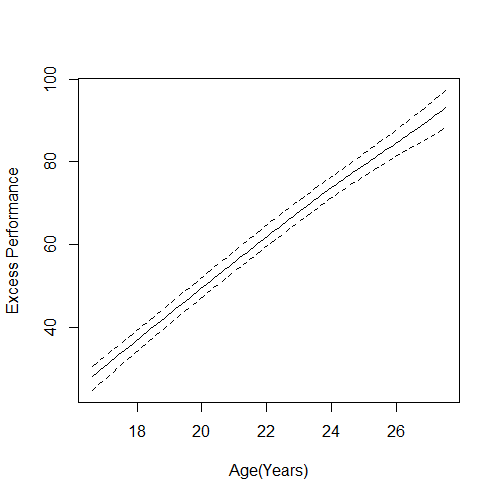}&
\includegraphics[scale=0.3]{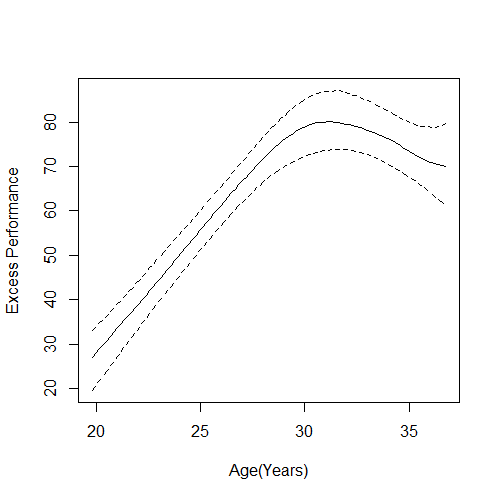}&
\includegraphics[scale=0.3]{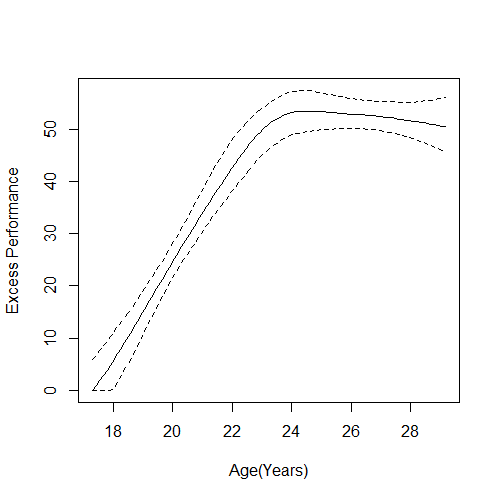}
\end{tabular}
\caption{\label{plt2} Individual and excess performance for Male Weightlifters}
\end{figure}

\begin{figure}[!htbp]
\centering
\begin{tabular}{c c c}
$\text{Athlete 10}$&$\text{Athlete 11}$&$\text{Athlete 12}$\\
\includegraphics[scale=0.3]
{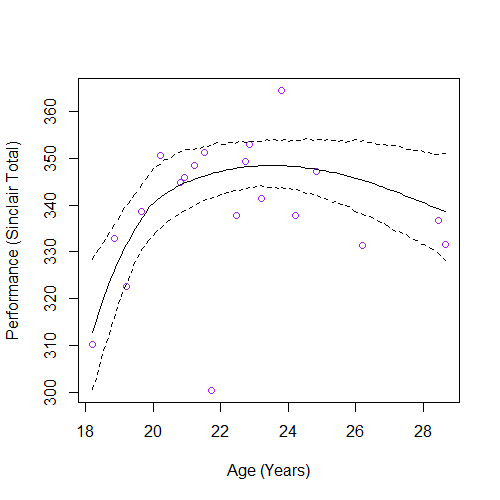}&
\includegraphics[scale=0.3]
{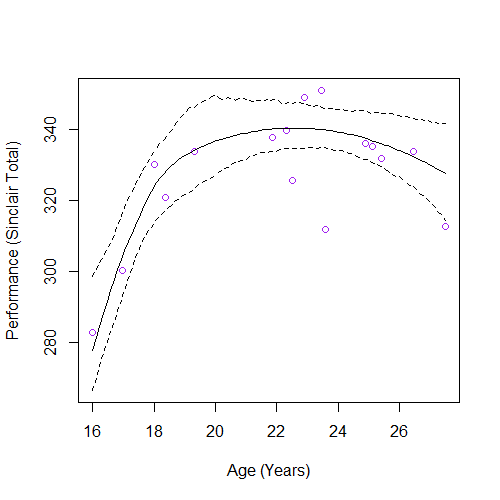}&
\includegraphics[scale=0.3]
{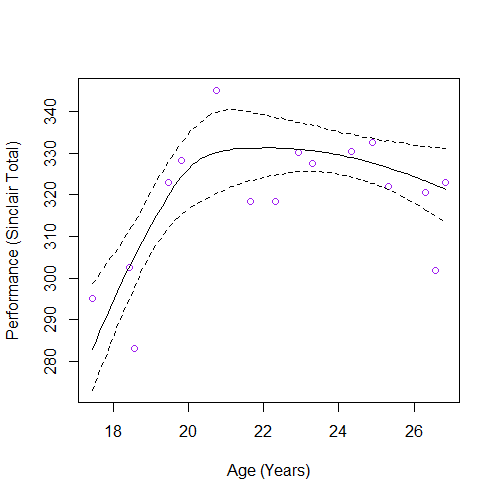}\\	
\includegraphics[scale=0.3]{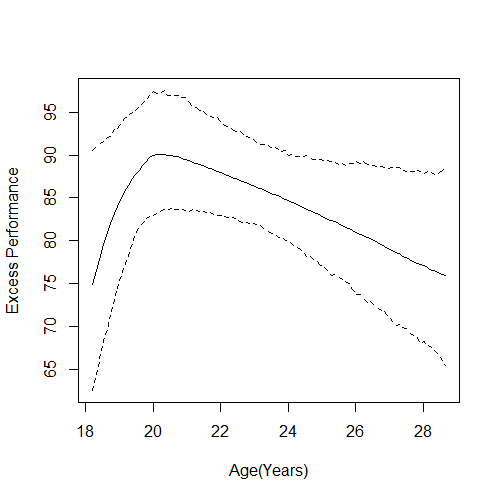}& \includegraphics[scale=0.3]{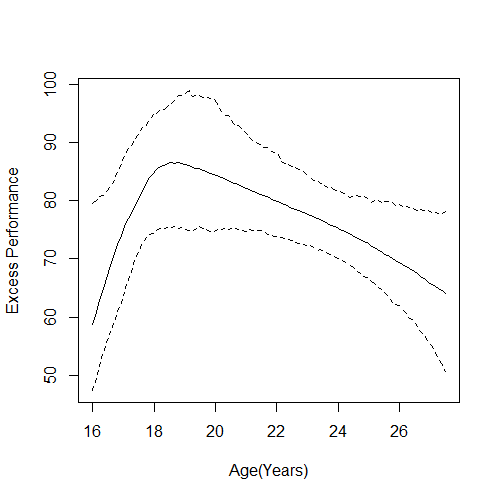}&
\includegraphics[scale=0.3]{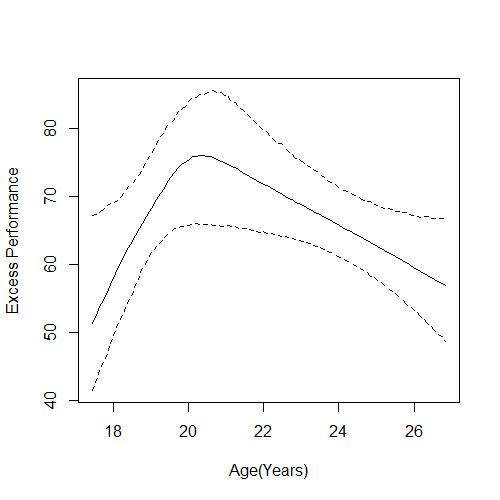}
\end{tabular}
\caption{\label{plt3} Individual and excess performance for Female Weightlifters}
\end{figure}

We make a comparison of the algorithm for the LME with skew $t$ errors to the results in \cite{griffin2022bayesian} using the same dataset of 100m and weightlifting for both genders. The computational time in table \ref{tab:tab1}. Although the skewness and degrees of freedom are not exactly the same between MCMC and EM, the estimates that they give are very close. Specifically, the skewness using EM for males and females is 0.7 and 0.9 respectively in contrast to 1.21 and 1.35 using MCMC for the 100m dataset. The degrees of freedom is 19.05 and 20.4 using EM whereas in MCMC is 18.4 and 19.6 respectively. 
For the weightlifting dataset we have -0.6 and -1.1 as estimates of skewness for males and females respectively with EM algorithm whereas for the MCMC equivalent we get -1.91 and -1.64 instead. Although we can observe a difference in estimates between the two methods both give a negative sign. Finally, the estimates for the degrees of freedom present some differences as well where EM's estimates for males and females are 17 and 5.8 with the MCMC's equivalent to be 7.8 and 6.9 respectively. \\
Finally, in the Figure \ref{plt1b} the population age performance is shown for both genders in 100m and weightlifting. In Figure  \ref{plt2a} the month effect and wind effect for the 100m are shown as well as the month effect for weightlifting being similar with the results of \cite{griffin2022bayesian}. Lastly, we present the performance and the individual-age performance plots for top three athletes where Figures \ref{plt1c}, \ref{plt1d} are about Males and Females of 100m dataset  and Figures \ref{plt2}, \ref{plt3} are about Males and Females of weightlifting.  

 
\section{Conclusions}\label{conclusions}

We develop an EM algorithm for LME with normal errors which is extended to an LME with skew $t$ errors using a variational Bayes approach. The latter approach could be extended to other non-normal error distributions using a latent variable representation which leads to a conditional normal linear model. A simulation study shows that the variational Bayes algorithm has good performance for a range of values of the degrees of freedom and skewness. An application of the algorithm to a longitudinal model used in the modelling of elite sporting performance show that the method has similar inference to MCMC on a real-world data set.

\section*{Acknowledgements}
This research was supported by a Partnership for Clean Competition research grant awarded to JH (Grant: 514).
Specialist and High Performance Computing systems were provided by Information Services at the University of Kent.

\bibliographystyle{chicago}
\bibliography{sample}

\newpage
\begin{appendices}
\section{EM calculations with normal errors}\label{EM1}

The full expression of  \eqref{analysed Q} is
\begin{align}\label{analysed Q2_append}
Q(\boldsymbol{\chi}) 
=&  \sum_{i=1}^M \sum_{k=1}^K w_{i,k}  \, \E_{\boldsymbol{\beta}_{i, k}, \sigma_{i, k}^2\mid \boldsymbol\gamma_{i, k},\boldsymbol{\chi}, \boldsymbol{y}_i}\left[ \log P\left(\boldsymbol{y}_i\mid \boldsymbol{\beta}_{i, k}, \sigma_{i, k}^2,\boldsymbol{\chi} \right)+ \log P\left(\boldsymbol{\beta}_{i, k}, \sigma_{i, k}^2\mid \boldsymbol\chi\right) \right] \notag\\
&+ \log P(\boldsymbol{\chi})\notag\\
 =&-\frac{1}{2}\sum_{i=1}^M \sum_{k=1}^K w_{i,k} \, \E_{\boldsymbol{\beta}_{i, k}, \sigma_{i, k}^2\mid \boldsymbol\gamma_{i, k},\boldsymbol{\chi}, \boldsymbol{y}_i}\left[
 \frac{1}{\sigma_{i, k}^2} \sum_{j=1}^{n_i} \left( y_{i, j} - \boldsymbol{X}_{i, j}\boldsymbol{\zeta}^*- \boldsymbol{S}_{i, j}\boldsymbol{\beta}_{i, k}\right)^2 + \log \sigma_{i, k}^2\right] \notag\\
&- \frac{1}{2}\sum_{i=1}^M \sum_{k=1}^K w_{i,k}\, \E_{\boldsymbol{\beta}_{i, k}, \sigma^2_{i, k}\mid \boldsymbol\gamma_{i, k},\boldsymbol{\chi}, \boldsymbol{y}_i} \left[\log(\psi\, \sigma_{i, k}^2) + p_{i, k}\,\log (g\,\sigma_{i, k}^2) + 
\frac{1}{\sigma_{i, k}^2}
(\boldsymbol{\beta}_{i, k})^T\,\boldsymbol{\Lambda}_{i, k}\,\boldsymbol{\beta}_{i, k}\right]\notag\\ 
& + M\,a\,\log b - M\,\log\Gamma(a) -(a + 1) \sum_{i=1}^M\sum_{k=1}^K w_{i,k}\, \E_{\boldsymbol{\beta}_{i, k}, \sigma_{i, k}^2\mid \boldsymbol\gamma_{i, k},\boldsymbol{\chi}, \boldsymbol{y}_i} \left[\log\sigma_{i, k}^2\right]\notag\\
&- b\sum_{i=1}^M \sum_{k=1}^K w_{i,k} \,\E_{\boldsymbol{\beta}_{i, k}, \sigma_{i, k}^2\mid \boldsymbol\gamma_{i, k},\boldsymbol{\chi}, \boldsymbol{y}_i} \left[\frac{1}{\sigma_{i, k}^2}\right] + M\left(\log\Gamma(a_1+b_1) -\log\Gamma\left(p + a_1 + b_1\right)\right)\notag\\
& - M\left(\log\Gamma(a_1) + \log\Gamma(b_1) \right) +\sum_{i=1}^M \sum_{k=1}^K w_{i, k}\,\left(\log \Gamma\left(p_{i, k}^{\gamma}+ a_1\right) + \log\Gamma\left(p - p_{i, k}^{\gamma} + b_1\right)\right)\notag\\
& -2\log \psi -\frac{1}{\psi} - \frac{1}{2} \log g - \log(1 + g)\notag
\end{align}
where $\boldsymbol{\Lambda}_{i, k} =  \mbox{diag}\left(\psi^{-1}, \underbrace{g^{-1}, \dots, g^{-1}}_{p_{i, k}^{\gamma} times}\right)$.

To work out the expectations, it's useful to note that if $\boldsymbol{X}\sim \N(\boldsymbol{\mu}, \boldsymbol{\Sigma})$, then 
\[
\E[\boldsymbol{X}] = \boldsymbol{\mu}, \qquad
\E[\boldsymbol{X}^T \boldsymbol{B} \boldsymbol{X} ]  =\mbox{tr}(\boldsymbol{B}\boldsymbol{\Sigma}^T) + \boldsymbol{\mu}^T \boldsymbol{B} \boldsymbol{\mu}\\
\]

\section{EM calculations with skew $t$ errors}\label{EM2}

The full expression of  \eqref{analysed Q2} is
\begin{align*}
Q(\boldsymbol{\chi}) 
=&  \sum_{i=1}^M \sum_{k=1}^K w_{i,k} 
\, \E_{\boldsymbol\psi_i} \left[\E_{\boldsymbol\phi_i} \left[ \log P\left(\boldsymbol{y}_i\mid \boldsymbol{\nu}_i,\boldsymbol{\chi} \right)+ \log P(\boldsymbol{\nu}_i) \right] \right]+ \log P(\boldsymbol{\chi})
\\
 =& -\sum_{i=1}^M 
 \sum_{k=1}^K \frac{w_{i,k}}{2} \,
 \E_{\boldsymbol\psi_i} \E_{\boldsymbol\phi_i}\left[
\frac{1}{\sigma_{i, k}^2} \sum_{j=1}^{n_i} \rho_{i, j} \left( \sqrt{1+c^2}\left(y_{i, j} - \boldsymbol{X}_{i, j}\boldsymbol{\zeta}^*- \boldsymbol{S}_{i, j}\boldsymbol{\beta}_i\right)- c\, d_{i, j}\right)^2\right]\\
&-\sum_{i=1}^M 
\sum_{k=1}^K\frac{ w_{i,k}}{2} \,
\E_{\boldsymbol\psi_i}  \E_{\boldsymbol\phi_i} \left[\sum_{j=1}^{n_i} \left( \log \left( \frac{\sigma_{i, k}^2}{(1+c^2)\rho_{i,j}}\right)+ \log \left( \frac{\sigma_{i, k}^2}{\rho_{i,j}}\right)\right) +\frac{\boldsymbol{\rho}_{i}^T\boldsymbol{d}_{i}^{2}}{\sigma^2_{i, k}}\right]\\
&- \sum_{i=1}^M \sum_{k=1}^K\frac{ w_{i,k}}{2} \,\E_{\boldsymbol\phi_i} \left[\log(\psi\,\sigma_{i, k}^2) +\frac{\boldsymbol{\beta}_{i, k}^T\boldsymbol{\Lambda}_i\boldsymbol{\beta}_{i, k}}{\sigma_{i, k}^2} + \frac{p_{i, k}}{2}\log (g^2\,\sigma_{i, k}^2)\right]\\
&+ M\left(a\log b - \log\Gamma(a)\right) -(a + 1) \sum_{i=1}^M\sum_{k=1}^K w_{i,k} \, \E_{\boldsymbol\phi_i}\left[\log\sigma^2_{i, k}- b\, \frac{1}{\sigma^2_{i, k}}\right]\\
&+ M\left(\log\Gamma(a_1+b_1) -\log\Gamma\left(p + a_1 + b_1\right) - \log\Gamma(a_1) - \log\Gamma(b_1) \right)\\
&+\sum_{i=1}^M \sum_{k=1}^K w_{i, k}\,\left(\log \Gamma\left(p_{i, k} + a_1\right) + \log\Gamma\left(p - p_{i, k} + b_1\right)\right)\\
&+\left( \frac{f}{2}\log\left(\frac{f}{2}\right)
- \log \Gamma\left(\frac{f}{2}\right)\right)\sum_{i=1}^M n_i
  + \sum_{i=1}^M\sum_{j=1}^{n_i} \left(
  \frac{f-2}{2}
  \E_{\boldsymbol\psi_i}\left[\log \rho_{i,j}\right]-\frac{f}{2} \E_{\boldsymbol\psi_i} \left[\rho_{i,j}\right]\right)\notag\\
  &-2\log \psi -\frac{1}{\psi} - \frac{1}{2}\log g - \log(1 + g)
  -\frac{1}{2}\frac{c^2}{100^2} +\log f -0.1f 
\end{align*}
where $\boldsymbol{\Lambda}_i =  \mbox{diag}\left(\psi^{-1}, \underbrace{(g^2)^{-1}, \dots, (g^2)^{-1}}_{p_i^{\gamma} times}\right)$.\\

It is useful to define 
$
X^{\star}_{i, j, m} = \sqrt{\E_{\boldsymbol\psi_i}[\rho_{i, j}]\,(1+c^2)} \, X_{i, j, m}
$ and $r_{i, j} =  \sqrt{E_{\boldsymbol\psi_i}[\rho_{i, j}]}\,\sqrt{1 + c^2} \,y_{i, j} - c\,
E_{\boldsymbol\psi_i}[\rho_{i, j}\, d_{i, j}]$ for $i = 1,\dots, M$, $j = 1, \dots, n_i$ and $m = 1, \dots, q$ and $S_{i, j, m} = \sqrt{
\E_{\boldsymbol\psi_i}[\rho_{i, j}]\,(1 + c^2)}\, S_{i, j, m}$ for $i = 1,\dots, M$, $j = 1, \dots, n_i$ and $m = 1, \dots, p_{i, k}$, and $\boldsymbol{r}_i = (r_{i, 1}, \dots, r_{i, n_i})$

The algorithm uses the following updates
\[
\boldsymbol\zeta^{\star} = \left(
\sum_{i=1}^M  
\boldsymbol{X}^{\star\,T}_{i} \boldsymbol{X}^{\star}_{i} \sum_{k=1}^K w_{i, k} \, \E_{\boldsymbol\phi_i}\left[ \frac{1}{\sigma_{i, k}^2}\right] \right)^{-1}
\left(\sum_{i=1}^M \boldsymbol{X}^{\star\,T}_{i} \sum_{k=1}^K w_{i, k}\, \left(\E_{\boldsymbol\phi_i}\left[\frac{1}{\sigma_{i, k}^2}\right]\boldsymbol{r}_{i} -  \E_{\boldsymbol\phi_i}\left[\frac{\boldsymbol{S}^{\star}_{i, k}\boldsymbol{\beta}_{i, k}}{\sigma_{i, k}^2} \right] \right)
\right) 
\]
and
\[
\psi = \frac{1}{M+4}\left( \sum_{i=1}^M
\sum_{k=1}^K w_{i, k}\,
\E\left[
\frac{\beta_{i,k,1}^2}{\sigma_{i, k}^2}\right] +2\right), 
\]

To find the maximizers of $a$ and $b$, we solve the following equations:
\[
\frac{\Gamma'(a)}{\Gamma(a)} =  \log b + \frac{1}{M}\sum_{i=1}^M \sum_{k=1}^K w_{i, k}\,\E\left[ \log\left(\frac{1}{\sigma_{i, k}^2}\right)\right],  \quad b = \frac{a\,M}{\sum_{i=1}^M\sum_{k=1}^K w_{i, k}\, \E\left[ \frac{1}{\sigma_{i,k}^2}\right]}
\]
In the same way, we update to $a_1$ to the maximizer of the equation
\[
 \log\Gamma(a_1+b_1) -\log\Gamma\left(p + a_1 + b_1\right)
- \log\Gamma(a_1)
 +\frac{1}{M}\sum_{i=1}^M \sum_{k=1}^K w_{i, k}\,\log \Gamma\left(p_{i, k} + a_1\right),
\]
 $b_1$ to the maximizer of the equation
\[
 \log\Gamma(a_1+b_1) -\log\Gamma\left(p + a_1 + b_1\right)
- \log\Gamma(b_1)
  +\frac{1}{M}\sum_{i=1}^M \sum_{k=1}^K w_{i, k}\, \log\Gamma\left(p - p_{i, k} + b_1\right),
\]
 $g$ to the maximizer of the equation
\[
 - \log g\sum_{i=1}^M \sum_{k=1}^K w_{i,k}\,  p_{i, k}
  - \frac{1}{g}\sum_{i=1}^M \sum_{k=1}^K w_{i,k}\, \E_{\boldsymbol{\beta}_{i, k}, \sigma^2_{i, k}\mid \boldsymbol\gamma_{i, k},\boldsymbol{\chi}, \boldsymbol{y}_i} \left[  \frac{\sum_{j=1}^{p_{i. k}}\boldsymbol{\beta}_{i, k, j}^2}{\sigma_{i, k}^2}\right] 
 -  \log g - 2\log(1 + g),
\]
 $c$ to the maximizer of the equation
\begin{align*}
 & -(1+c^2) \sum_{i=1}^M 
   \sum_{j=1}^{n_i}
 \E_{\boldsymbol\psi_i}\left[\rho_{i, j} \right] 
 \sum_{k=1}^K \frac{w_{i,k}}{2} 
  \E_{\boldsymbol\phi_i}\left[
\frac{1}{\sigma_{i, k}^2} 
\left(y_{i, j} - \boldsymbol{X}_{i, j}\boldsymbol{\zeta}^*- \boldsymbol{S}_{i, j}\boldsymbol{\beta}_i\right)^2\right]\\
 & +2c\sqrt{1+c^2}\sum_{i=1}^M \sum_{j=1}^{n_i}
 \E_{\boldsymbol\psi_i}\left[
 \rho_{i, j} \,d_{i, j}\right]
 \sum_{k=1}^K \frac{w_{i,k}}{2} \,
  \E_{\boldsymbol\phi_i}\left[
\frac{1}{\sigma_{i, k}^2}  \left(y_{i, j} - \boldsymbol{X}_{i, j}\boldsymbol{\zeta}^*- \boldsymbol{S}_{i, j}\boldsymbol{\beta}_i\right)\right]\\
& -c^2\sum_{i=1}^M   \sum_{j=1}^{n_i} \E_{\boldsymbol\psi_i}\left[\rho_{i, j}   \,d_{i, j}^2\right]
 \sum_{k=1}^K \frac{w_{i,k}}{2}
  \E_{\boldsymbol\phi_i}\left[
\frac{1}{\sigma_{i, k}^2} \right]
 + \log \left( 1+c^2\right) \sum_{i=1}^M n_i   -\frac{1}{2}\frac{c^2}{100^2} 
\end{align*}
and  $f$ to the maximizer of the equation
\begin{align*}
&+\left( \frac{f}{2}\log\left(\frac{f}{2}\right)
- \log \Gamma\left(\frac{f}{2}\right)\right)\sum_{i=1}^M n_i
  + \sum_{i=1}^M\sum_{j=1}^{n_i} \left(
  \frac{f-2}{2}
  \E_{\boldsymbol\psi_i}\left[\log \rho_{i,j}\right]-\frac{f}{2} \E_{\boldsymbol\psi_i} \left[\rho_{i,j}\right]\right)\notag\\
  & +\log f -0.1f.
\end{align*}

\section{Further simulation results}

\begin{figure}[h!]
\begin{center}
\begin{tabular}{cc}
$f = 5$, $c = 0$ &   $f = 5$, $c = 4$\\
\includegraphics[scale=0.4]{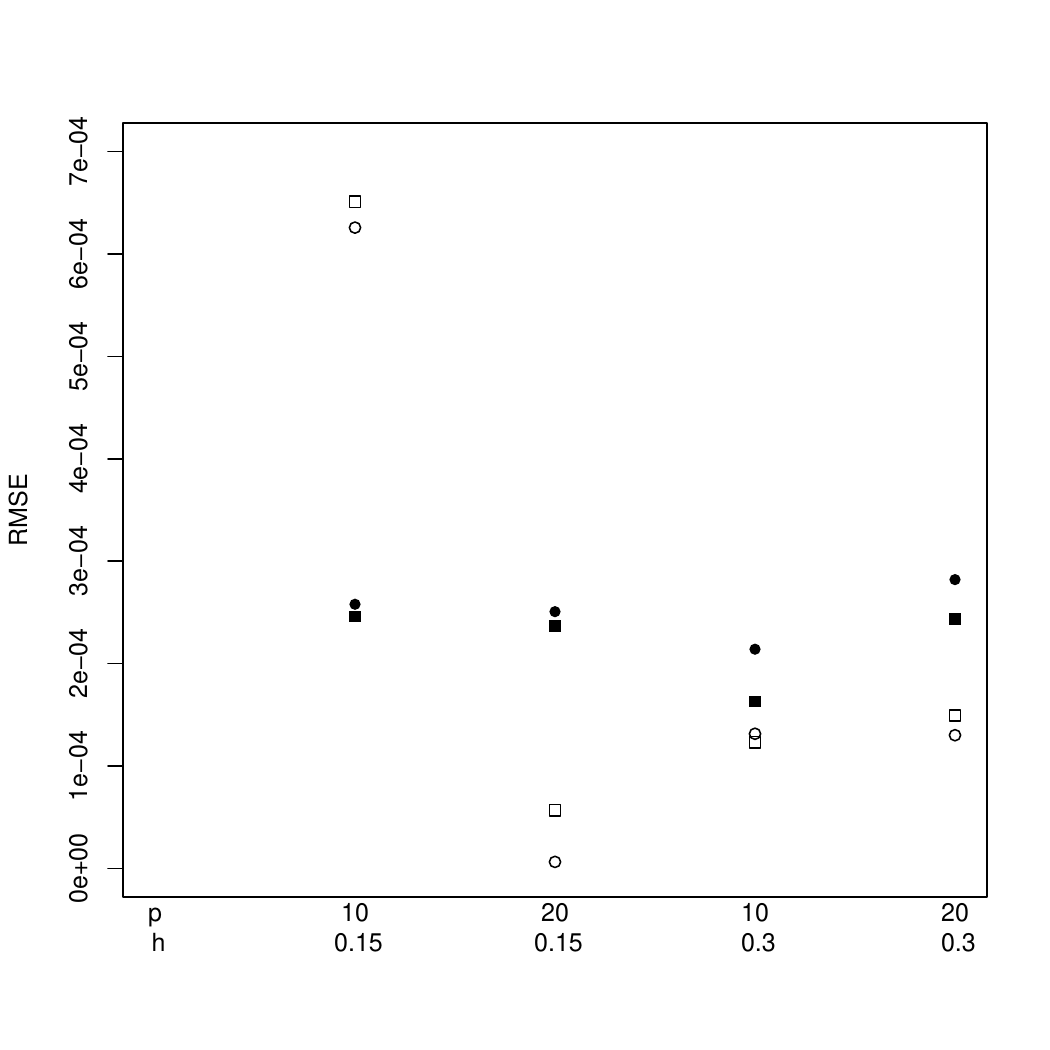} &
\includegraphics[scale=0.4]{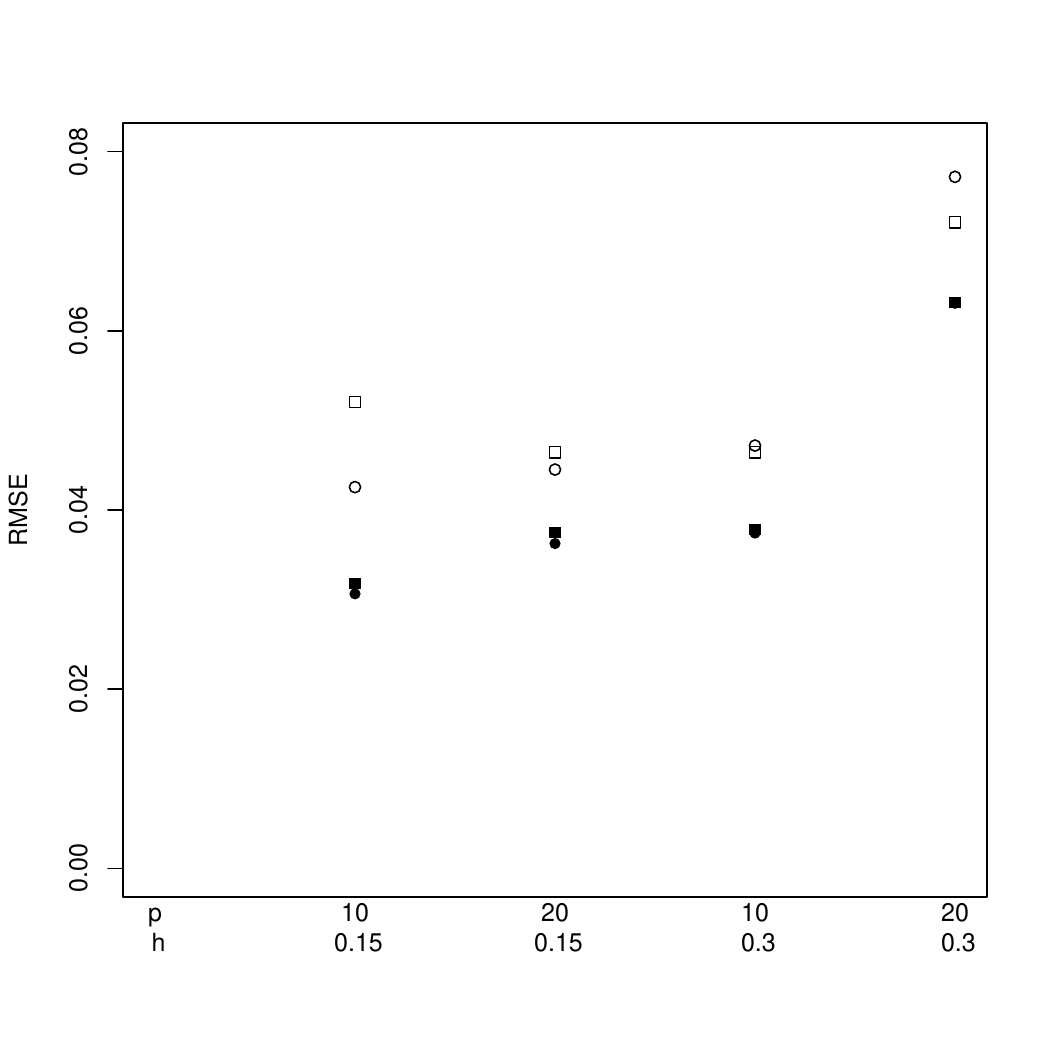}\\
$f = 20$, $c = 0$ &   $f = 20$, $c = 4$\\
\includegraphics[scale=0.4]{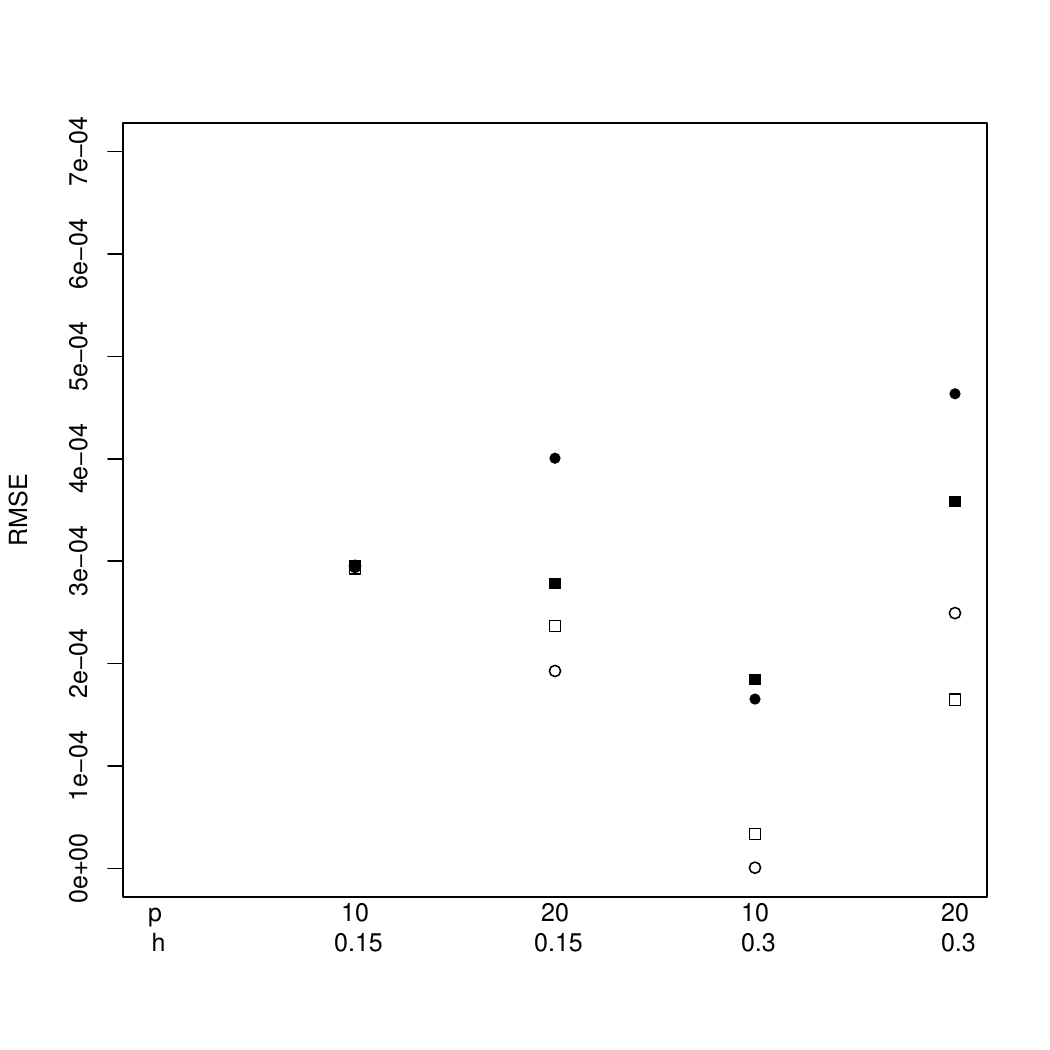} &
\includegraphics[scale=0.4]{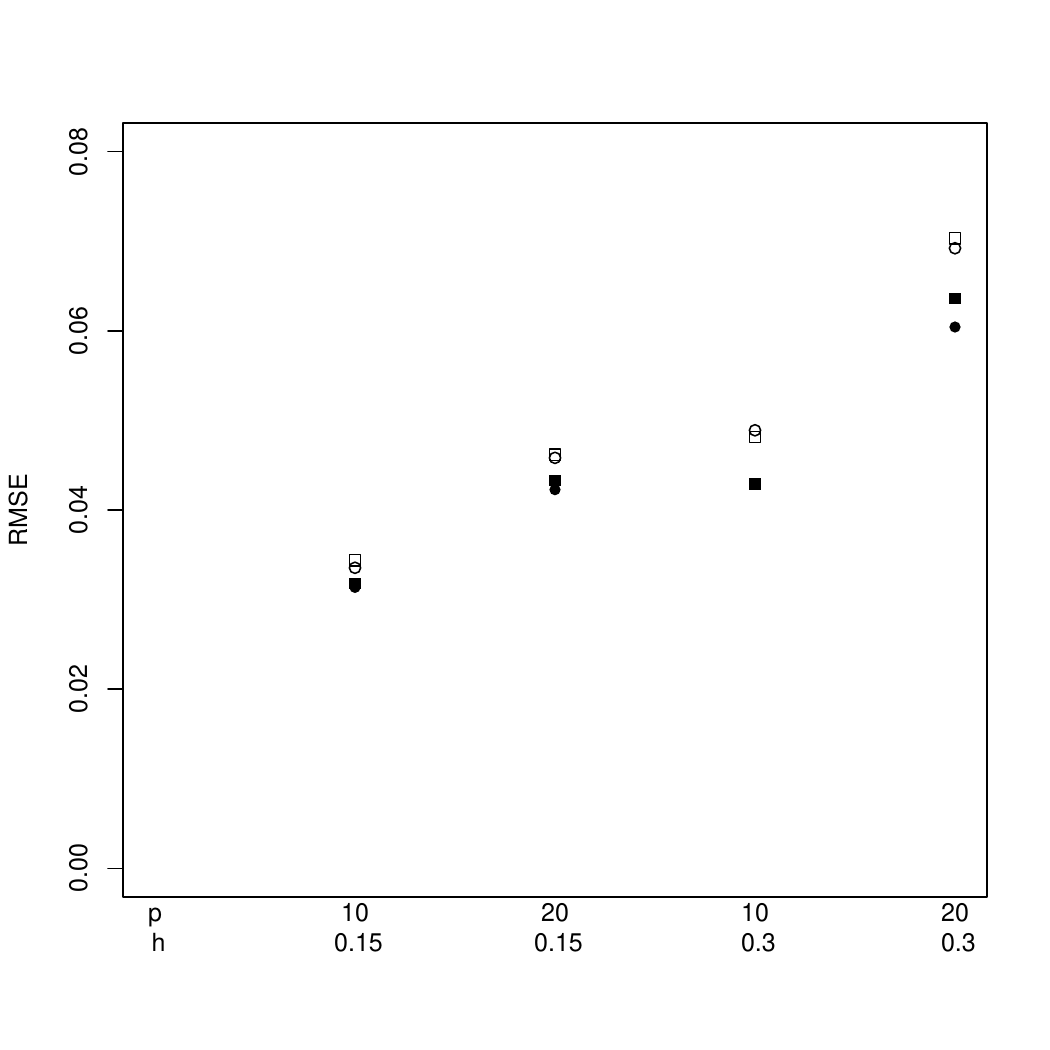}
\end{tabular}
\end{center}
\caption{Simulation study: RMSE for $c$. The symbols represent
 $q = 0.15$ and window = 30 (\raisebox{0.02cm}{\scalebox{0.6}{$\square$}}), 
 $q = 0.15$ and window = 100 ($\circ$),  
 $q = 0.3$ and window = 30 (\raisebox{0.02cm}{\scalebox{0.6}{$\blacksquare$}}), and 
 $q = 0.3$ and window = 100 ($\bullet$)
 }
\end{figure}

\begin{figure}[h!]
\begin{center}
\begin{tabular}{cc}
$f = 5$, $c = 0$ &   $f = 5$, $c = 4$\\
\includegraphics[scale=0.4]{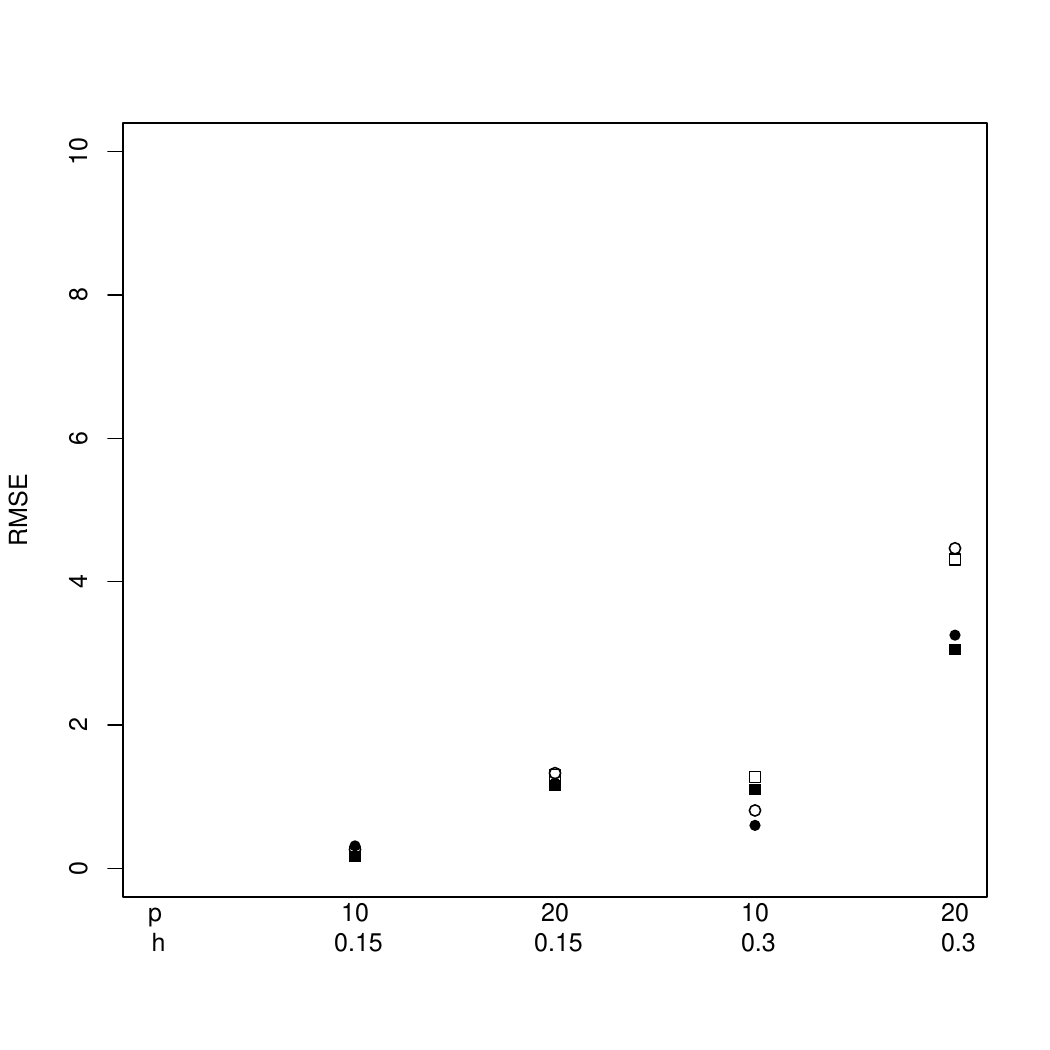} &
\includegraphics[scale=0.4]{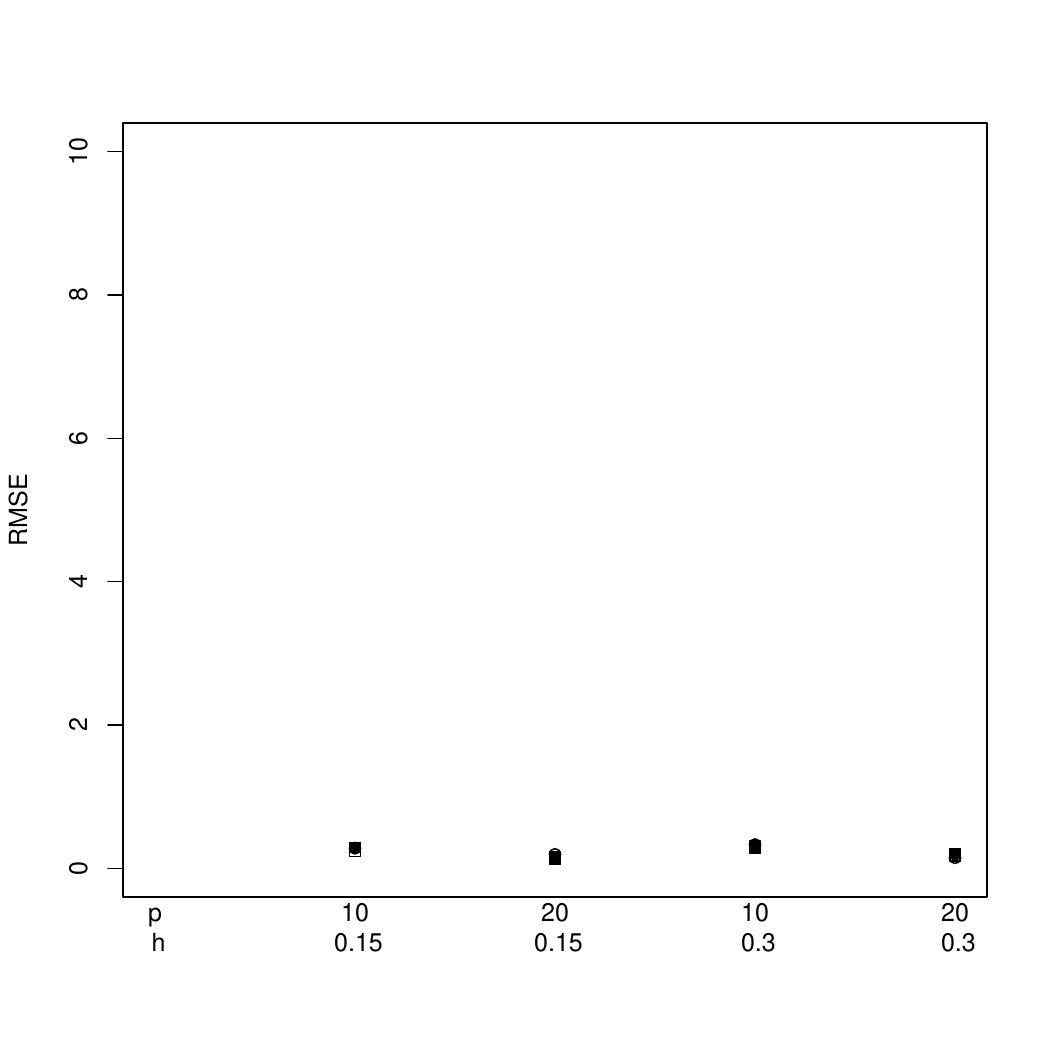}\\
$f = 20$, $c = 0$ &   $f = 20$, $c = 4$\\
\includegraphics[scale=0.4]{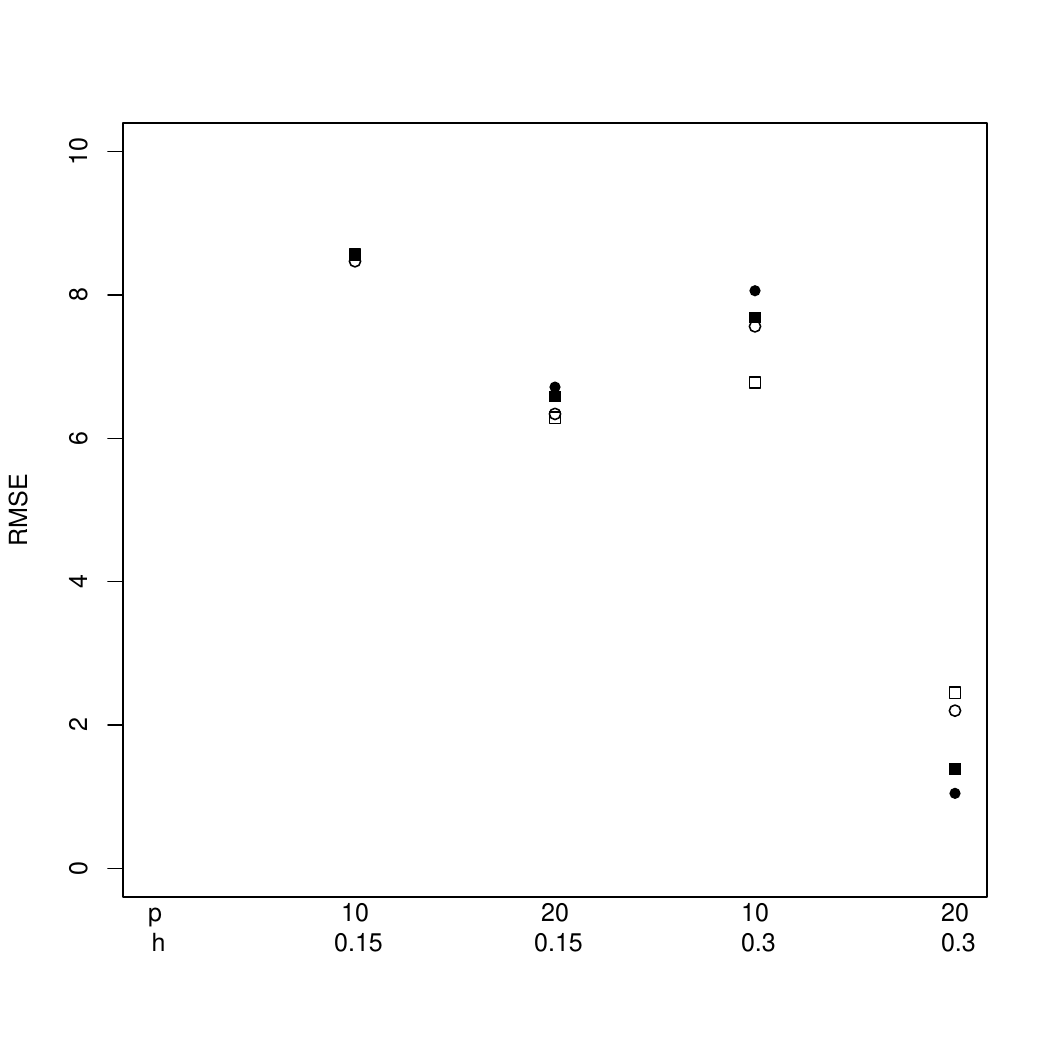} &
\includegraphics[scale=0.4]{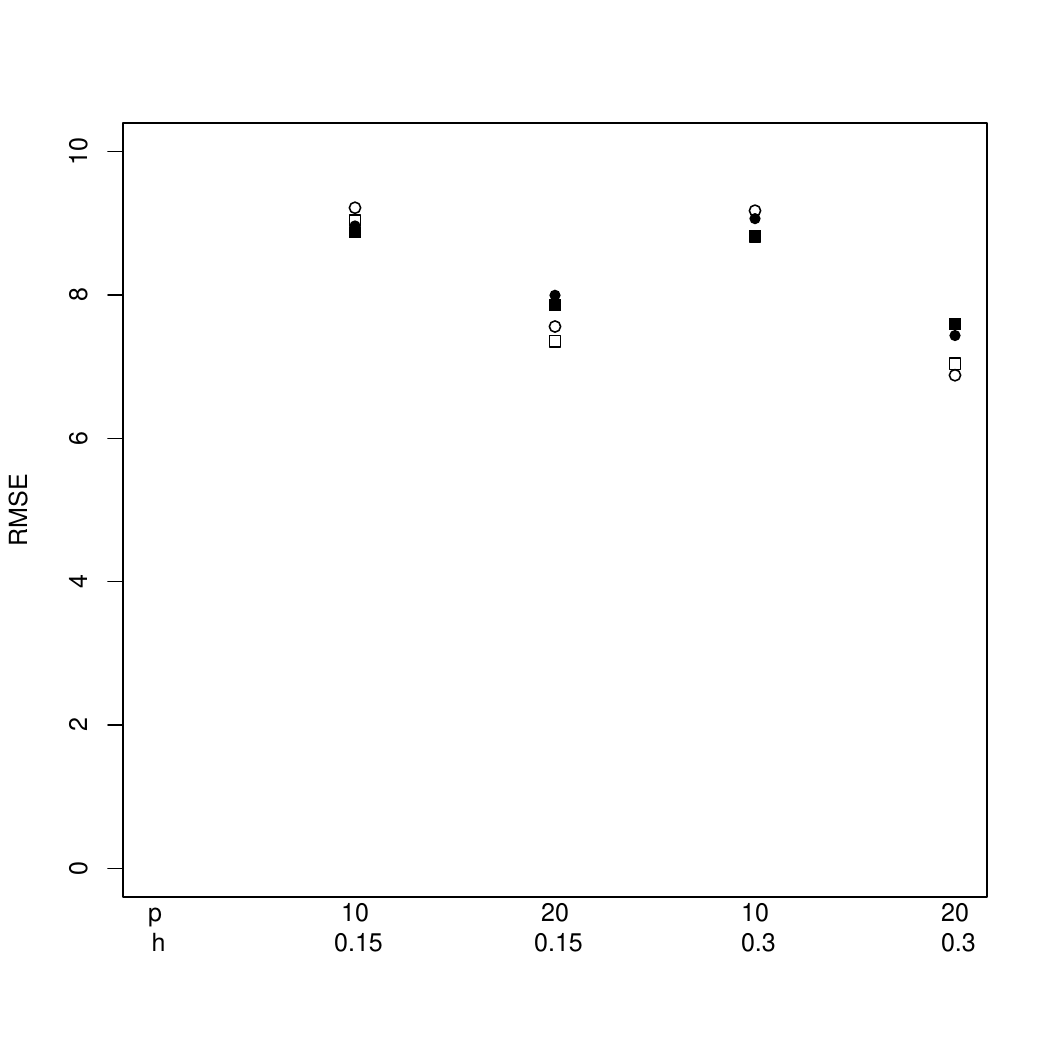}
\end{tabular}
\end{center}
\caption{Simulation study: RMSE for $f$. The symbols represent
 $q = 0.15$ and window = 30 (\raisebox{0.02cm}{\scalebox{0.6}{$\square$}}), 
 $q = 0.15$ and window = 100 ($\circ$),  
 $q = 0.3$ and window = 30 (\raisebox{0.02cm}{\scalebox{0.6}{$\blacksquare$}}), and 
 $q = 0.3$ and window = 100 ($\bullet$)
 }
\end{figure}

\end{appendices}
\end{document}